\documentclass[a4paper,11pt, twocoloumn]{article}
\usepackage{amsmath, amssymb}
\usepackage{array}
\usepackage[english]{babel}
\usepackage{fancyhdr}
\usepackage[T1]{fontenc}
\usepackage[margin=1in]{geometry}
\usepackage{graphicx}
\usepackage{siunitx}
\usepackage[utf8]{inputenc}
\usepackage{lastpage}
\usepackage{longtable}
\usepackage{mathtools, slashed}
\usepackage{multicol}
\usepackage{multirow}
\usepackage{subfigure}
\usepackage{tabulary}
\usepackage{tabularx}
\usepackage{verbatim}
\usepackage{enumitem}
\usepackage{capt-of}
\usepackage{xcolor}
\setlist[enumerate]{noitemsep}
\setlist[itemize]{noitemsep}
\setlist[description]{noitemsep}
\usepackage[margin=10pt, font=small, labelfont=bf]{caption}
\usepackage{longtable}
\usepackage{nameref}
\usepackage{hyperref}

\begin{document}
	
\renewcommand{\subsectionmark}[1]{\markboth{#1}{}}

\begin{center}
	{\Large 
		\textbf{
		Localisation-to-delocalisation transition of moir\'{e} excitons in WSe$_2$/MoSe$_2$ heterostructures
		}}
\end{center}

\vspace{1 mm}

\begin{center}
	{\large Elena Blundo,$^{1,*}$  Federico Tuzi,$^{1}$ Salvatore Cianci,$^{1}$ Marzia Cuccu,$^{1}$ Katarzyna Olkowska-Pucko,$^{2}$ \L{}ucja Kipczak,$^{2}$ Giorgio Contestabile,$^{1}$ Antonio Miriametro,$^{1}$ Marco Felici,$^{1}$ Giorgio Pettinari,$^{3}$ Takashi Taniguchi,$^{4}$ Kenji Watanabe,$^{5}$ Adam Babi\'{n}ski,$^{2}$ Maciej R. Molas,$^{2}$ and Antonio Polimeni$^{1,*}$}
\end{center}

\begin{center}
	\textit{\mbox{}$^1$ Physics Department, Sapienza University of Rome, 00185, Roma, Italy.\\
 \mbox{}$^2$ Institute of Experimental Physics, Faculty of Physics, University of Warsaw, Pasteura 5, 02-093 Warsaw, Poland\\
	\mbox{}$^3$ Institute for Photonics and Nanotechnologies (CNR-IFN), National Research Council, 00133, Rome, Italy\\
	\mbox{}$^4$ International Center for Materials Nanoarchitectonics, National Institute for Materials Science, 1-1 Namiki, Tsukuba 305-0044, Japan.\\
	\mbox{}$^5$ Research Center for Functional Materials, National Institute for Materials Science, 1-1 Namiki, Tsukuba 305-0044, Japan\\
	}
\end{center}

\mbox{}$^*$ Corresponding authors: antonio.polimeni@uniroma1.it, elena.blundo@uniroma1.it

\begin{center}
    24 April 2023
\end{center}

\date{}

\vspace{10 mm}

\noindent
\textbf{Abstract}\\
\mbox{}\\
Moir\'{e} excitons (MXs) are electron-hole pairs localised by the periodic (moir\'{e}) potential forming in two-dimensional heterostructures (HSs). MXs can be exploited, \emph{e.g.}, for creating nanoscale-ordered quantum emitters and achieving or probing strongly correlated electronic phases at relatively high temperatures. Here, we studied the exciton properties of a WSe$_2$/MoSe$_2$ HS from \emph{T}=6 K to room temperature using time-resolved and continuous-wave micro-photoluminescence, also under magnetic field. The exciton dynamics and emission lineshape evolution with temperature show clear signatures that MXs de-trap from the moir\'{e} potential and turn into free interlayer excitons (IXs) at $T\gtrsim$120 K. The MX-to-IX transition is also apparent from the exciton magnetic moment reversing its sign when the moir\'{e} potential is not capable to localise excitons at elevated temperatures. Concomitantly, the exciton formation and decay times reduce drastically. Thus, our findings establish the conditions for a truly confined nature of the exciton states in a moir\'{e} superlattice with increasing temperature.



\twocolumn

\section*{Introduction}
Two-dimensional (2D) heterostructures (HSs) can be formed by stacking two (or more) monolayers (MLs) of different van der Waals crystals. 2D HSs offer a countless number of combinations thanks to the nearly arbitrary choice of the chemical composition of the individual constituents and the control of their relative angular alignment \cite{moire_review}. Inherent to the stacking process is the formation of a moir\'{e} superlattice that superimposes on the topographic and electronic structure of the single MLs. This phenomenon has been particularly investigated in HSs made of transition metal dichalcogenide (TMD) semiconductors, which feature a sizeable band gap \cite{zhang_IX_couplings,YU_quantum_emitters,Tran:2019jl,Seyler_moire_excitons,Choi_IX_lifetime,LI_IX_transport,moire_photonics_review}. The moir\'{e} potential can be as deep as 100 meV \cite{Tran:2019jl,LI_IX_transport} and can localise both intralayer excitons (Xs) residing in the MLs of the HS \cite{HS_Intra_Zhang} and interlayer excitons (IXs) \cite{YU_quantum_emitters,Tran:2019jl,Seyler_moire_excitons,Choi_IX_lifetime} and trions \cite{Liu_moire_lattice_period}, in which different charge carriers reside in the different layers of the HS. Moir\'{e}-confined IXs (hereafter, moir\'{e} excitons, MXs) are especially interesting as they can be exploited as nanoscale ordered arrays of quantum emitters \cite{Baek:2020cf,YU_quantum_emitters}. Furthermore, their space-indirect character endows IXs, and specifically MXs, with long lifetimes \cite{Tran:2019jl,Seyler_moire_excitons} that, in conjunction with the depth of the moir\'{e} potential \cite{Tran:2019jl,LI_IX_transport}, make them suitable for the observation of high-temperature ($>100$ K) Bose-Einstein condensates, as shown in a WSe$_2$/MoSe$_2$ HS \cite{Zang_BE_IXcondensation}. The topology of the moir\'{e} potential also induces strongly correlated electron and exciton states \cite{Wang_moire_mott,correlated_exciton_review} that led to the observation of an exciton insulator surviving up to 90 K in a WS$_2$/bilayer-WSe$_2$ HS \cite{exciton_insulator}. In addition, the MXs themselves were employed as a probe of the existence of Mott insulators and Wigner crystals in WSe$_2$/WS$_2$ HSs at relatively large temperatures \cite{Mott_moire_Regan,correlated_highT}.

For boson condensates and highly correlated charge systems, as well as quantum photonics applications, the thermal stability of the moir\'{e}-induced confinement of the excitons plays a crucial role and a fundamental question arises: \emph{Up to which extent can MXs be regarded as truly moir\'{e}-confined?} We addressed this important aspect by investigating the effect of the lattice temperature and of the photogenerated exciton density on the localisation of MXs as resulting from their: (i) luminescence intensity and lineshape, (ii) temporal dynamics, (iii) magnetic moments. Specifically, we studied the emission properties of an exemplary WSe$_2$/MoSe$_2$ HS by continuous-wave (cw) micro-photoluminescence ($\mu$-PL) measurements, also under magnetic field, and by time-resolved (tr) $\mu$-PL. Low-temperature ($T$=6 K) tr-$\mu$-PL shows that the MX signal is characterised by different spectral components, with formation and recombination dynamics indicative of the presence of a multi-level electronic potential \cite{Wu_theory_moire_potential,Tran:2019jl,Choi_IX_lifetime,time_resolved_novoselov}. The temperature evolution of the HS emission properties presents clear signatures of IX de-trapping from the moir\'{e} potential at $T\approx120$ K and the ensuing spectral predominance of free IXs at higher temperatures. Concomitantly, Zeeman-splitting measurements reveal an unexpected sign reversal of the exciton magnetic moment taking place with the temperature-induced MX transition to a free IX regime. This transition is paralleled by a strong reduction of both the emission rise and decay times, which mirrors the faster formation and recombination dynamics, respectively, of the free IXs.

\section*{Results and discussion}

\noindent
\textbf{Moir\'{e} exciton dynamics at low temperature}\\
The investigated WSe$_2$/MoSe$_2$ HS was fabricated by first depositing a MoSe$_2$ flake containing a ML on a Si/SiO$_2$ substrate and then depositing a WSe$_2$ ML on top; see Methods for other details.
\begin{figure}[htbp]
\includegraphics[width=0.45\textwidth]{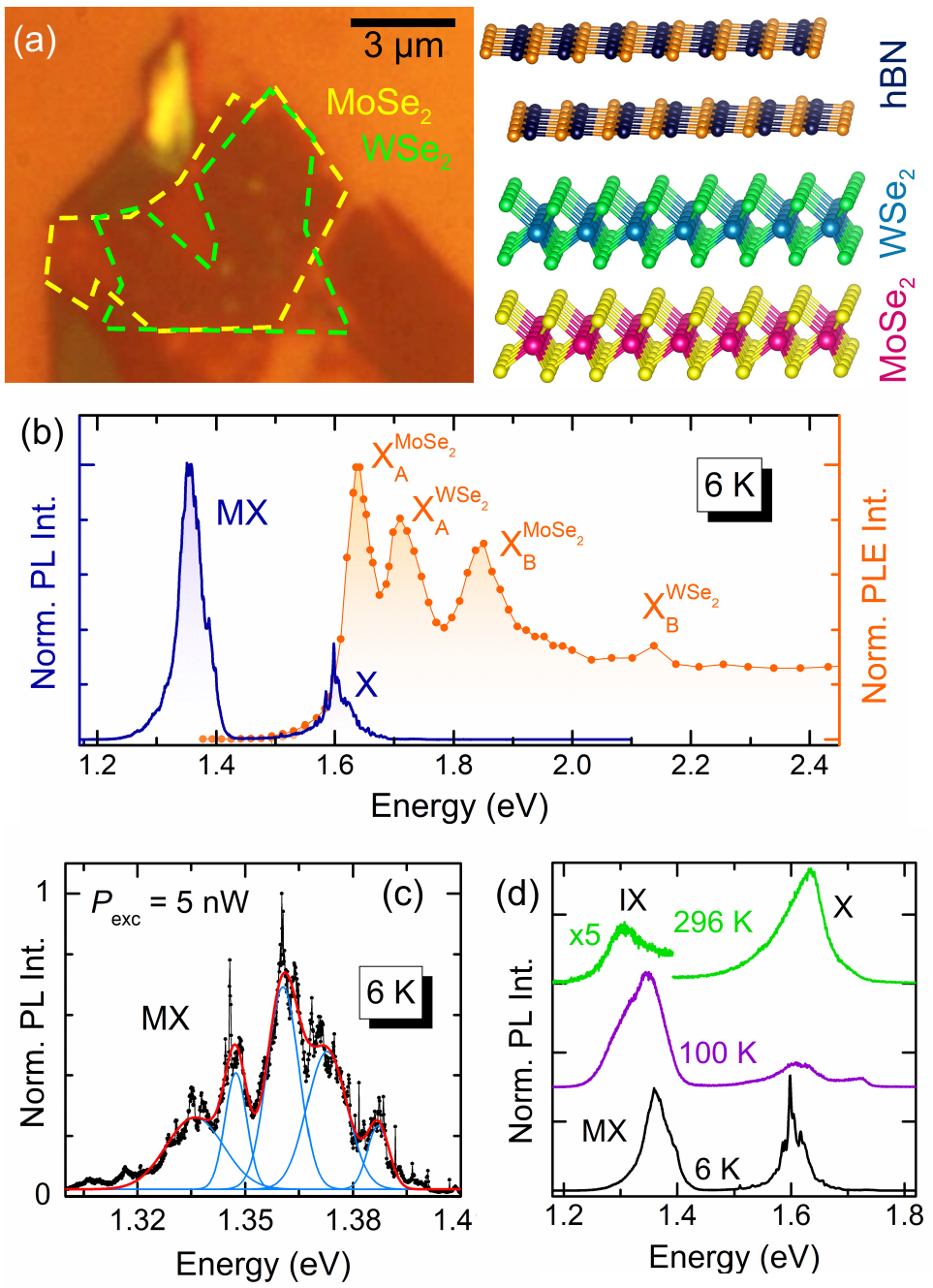}
\caption{\textbf{Optical properties of the WSe$_2$/MoSe$_2$ R-type HS} (a) Optical micro-graph (left) and sketch (right) of the investigated HS with the MoSe$_2$ ML being the layer closest to the Si/SiO$_2$ substrate and the h-BN layer protecting the HS from the ambient environment. (b) Low-\emph{T} $\mu$-PL and $\mu$-PLE spectra of the HS, left and right axis, respectively. In the $\mu$-PL spectrum ($P_\mathrm{exc} = 2 ~\mu$W), X indicates the intralayer exciton recombination from localised states of the MoSe$_2$ and WSe$_2$ monolayers (lower- and higher-energy side, respectively). MX is the moir\'{e} exciton band. In the $\mu$-PLE spectrum, four exciton resonances are observed. These resonances can be attributed to the A and B excitons (where the hole sits in the upper, A, and lower, B, spin-split valence band maximum at K, and the electron sits in the spin-split conduction band minimum at K with same spin) of the MoSe$_2$ and WSe$_2$ layers. (c) $\mu$-PL spectrum of the MX band acquired with very low laser power excitation (5 nW). The spectrum can be reproduced by five Gaussian functions (azure: single components; red line: total fit) that are spaced by $(12.8 \pm 1.3)$ meV. The very narrow lines that make up the broader Gaussian peaks correspond to single MXs recombining in moir\'{e} minima. (d) $\mu$-PL spectra recorded at different temperatures (and $P_\mathrm{exc} = 20 ~\mu$W). The moir\'{e}/interlayer (MX/IX) exciton band  is visible up to room temperature. X indicates the exciton band related to the single layer MoSe$_2$ and WSe$_2$ constituents of the HS.}
\label{fig:1}
\end{figure}
The HS was then capped with a thin hexagonal boron nitride (h-BN) layer to prevent oxidation. The relative twist angle between the two MLs is $ \theta\approx 0^\circ$ (R-type HS) as discussed next.
Fig.\ \ref{fig:1}(a) shows an optical microscope image of the HS along with its sketch. Cw and tr-$\mu$-PL measurements were carried out at variable laser excitation power $P_\mathrm{exc}$ and temperature $T$ using a confocal microscope setup. For $\mu$-PL excitation ($\mu$-PLE) measurements, we employed the same setup using a wavelength-tunable laser as excitation source. Magneto-$\mu$-PL measurements were performed at variable temperature in a superconducting magnet up to 12 T, with the field perpendicular to the HS plane. Further details are reported in the Methods section.


Fig.\ \ref{fig:1}(b) shows the $T$=6 K $\mu$-PL spectrum (blue line) of the investigated WSe$_2$/MoSe$_2$ HS. Two bands are observed. The one peaked at 1.6 eV, labelled X, is due to a group of localised (intralayer) exciton states originating from the MoSe$_2$ ML with a small contribution from similar transitions in the WSe$_2$ ML on the higher energy side of the band. The band centred at $\approx1.36$ eV, labelled MX, is due to MX recombination (with the electron and hole being confined in the MoSe$_2$ and WSe$_2$ layer, respectively), as also reported in other works \cite{Tran:2019jl,Seyler_moire_excitons,Miller_IX_decay_time,Choi_IX_lifetime,time_resolved_novoselov}. The orange line in Fig.\ \ref{fig:1}(b) is the $\mu$-PLE spectrum obtained by monitoring the MX signal while scanning the excitation laser wavelength. The MX signal shows a resonant contribution from the MoSe$_2$ and WSe$_2$ ML exciton states of the HS, thus confirming the interlayer nature of the MX band. We point out that, at variance with Ref.\ \cite{Barre_IX_absorption}, no MX-related absorption feature is instead observed in the $\mu$-PLE data due to the much smaller oscillator strength of the MX absorption. 

Fig.\ \ref{fig:1}(c) displays the MX spectrum recorded at $T$=6 K with $P_\mathrm{exc}$=5 nW (corresponding to 0.64 W/cm$^2$). The spectrum can be deconvoluted into several Gaussian components. The latter are equally spaced by $(12.8 \pm 1.3)$ meV, reflecting the quantised states of the moir\'{e} potential \cite{Tran:2019jl,Choi_IX_lifetime,Wu_theory_moire_potential,time_resolved_novoselov}. The Gaussian lineshape maps onto the ensemble of MXs confined in randomly distributed moir\'{e} minima due to the inevitable imperfections present in the HS plane. The very narrow lines superimposed on the multi-gaussian lineshape of the MX band likely correspond to single MXs confined in just one moir\'{e} minimum \cite{brotons_prx,LI_IX_transport}. The centroid energy of the MX band (1.357 eV) indicates that the investigated HS is R-type ($\theta\approx 0^\circ$) \cite{Seyler_moire_excitons,Choi_IX_lifetime,Tran:2019jl,Wang_moire_mott,Li_localized_IX}. In fact, for H-type HSs ($\theta\approx 60^\circ$) the MX recombination is centred at a higher energy ---by about 40 meV \cite{Seyler_moire_excitons,Liu_moire_lattice_period,Nagler_magnetoPL,Wang_singlet_triplet,Li_IX_PL_saturation,Zhang_singlet_triplet}--- due to the shallower moir\'{e} potential for H-type with respect to R-type HSs \cite{LI_IX_transport}. From the spacing between the MX states, as detailed in \textcolor{purple}{Supporting Note 1}, we estimate a moir\'{e} superlattice period $a_\mathrm{m}$ of about 40 nm, which corresponds to $ \theta$ of about half a degree \cite{Liu_moire_lattice_period}. From the HS period, we deduce that about 600 moir\'{e} minima are probed within the laser spot (radius equal to $\approx 500$ nm). The excellent alignment leads to a sizeable signal of the HS IXs up to room temperature, as shown in Fig.\ \ref{fig:1}(d). Note that the recombination from the HS is indicated as MX at $T$=6 K and as IX at $T$=296 K, qualitatively hinting at a temperature-induced transition in the character of the exciton. We investigated such transition by studying the temporal evolution of the HS exciton signal, its dependence on the number of photogenerated carriers and by determining the exciton gyromagnetic factor at different temperatures.

%
\begin{figure*}[htpb]
\includegraphics[width=1.0\textwidth]{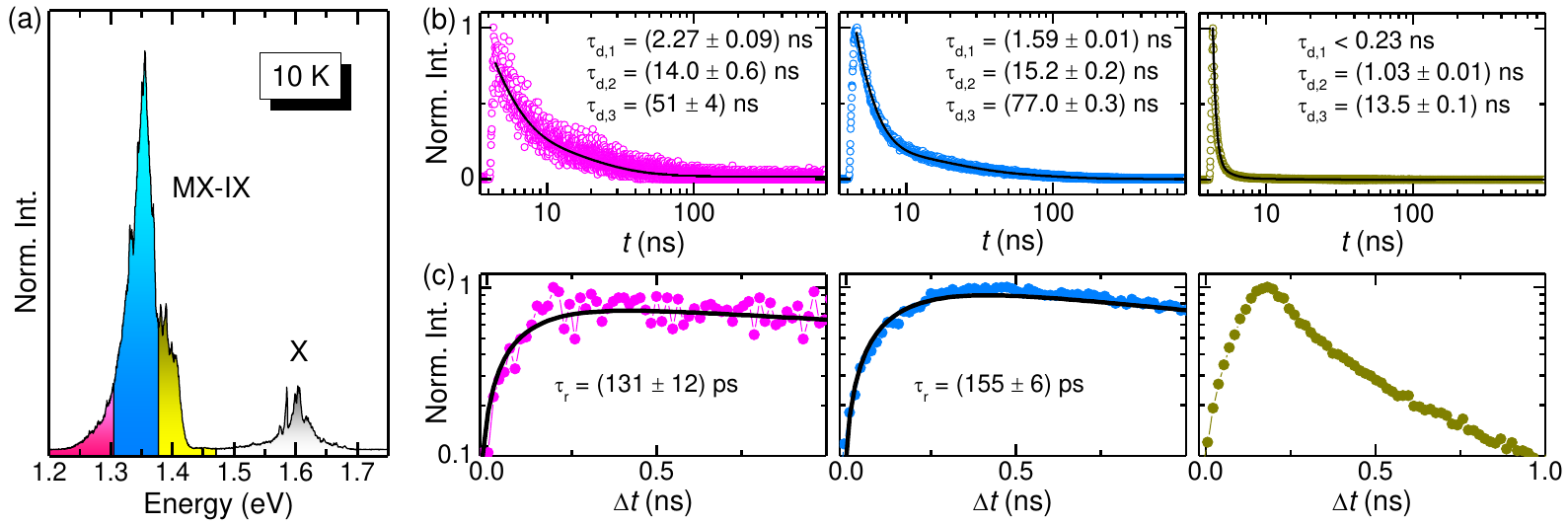}
\caption{\textbf{Decay and rise of the moir\'{e} exciton band}. (a) $T$=10 K (and $P_\mathrm{exc}$=1 $\mu$W) $\mu$-PL spectrum of the investigated WSe$_2$/MoSe$_2$ HS. MX-IX indicates the moir\'{e}/interlayer exciton, and X indicates the intralayer exciton recombination. Three different spectral regions are highlighted on the MX-IX band (MX-IX is meant to indicate the mixed character of the 1.4 eV component). On each of these regions, the $\mu$-PL time evolution was recorded. (b) Time-evolution of the $\mu$-PL signal recorded in the $\Delta$t=0-800 ns interval from the laser pulse on the three spectral regions highlighted in panel (a) (note also the colour code). The decay time $\tau_\mathrm{d,n}$ values obtained by fitting the data via Eq.\ \ref{eq:decay} (see solid lines) are displayed. (c) The same as (b) for $\Delta$t=0-1.0 ns. The rise time $\tau_\mathrm{r}$ values displayed in the panels are those used to reproduce the data with Eq.\ \ref{eq:rise} (see solid lines). The data in the right-most panel are close to the resolution limit and could not be fitted reliably.}
\label{fig:2}
\end{figure*}
We first describe the tr-$\mu$-PL results at $T$=10 K, where most of the HS emission is due to the MX recombination. Fig.\ \ref{fig:2}(a) shows the $\mu$-PL spectrum of the investigated WSe$_2$/MoSe$_2$ HS recorded at a power 200 times larger ($P_\mathrm{exc}$=1 $\mu$W, \emph{i.e.}, 128 W/cm$^2$) than in Fig.\ \ref{fig:1}(c). This results in a non negligible contribution from a component centred at about 1.4 eV, which can be assigned (totally or partly) to free (or moir\'{e}-de-trapped) IXs. Three different spectral windows are highlighted in Fig.\ \ref{fig:2}(a). For each of them, panel (b) and panel (c) display the corresponding $\mu$-PL signal time evolution from the laser pulse up to 800 ns and in the time interval (0-1) ns, respectively. In the former range, the decay part of the data can be fitted by
\begin{equation}
    I_\mathrm{decay}(t) = \sum_{ \mathrm{n} = 1}^{3} A_\mathrm{d,n} \cdot \exp{\left( -\frac{t-t_0}{\tau_\mathrm{d,n}} \right),}
    \label{eq:decay}
\end{equation}
where $t_0$ is a reference time, $\tau_\mathrm{d,n}$ is the decay time relative to the $n$-th component, whose weight is given by $w_\mathrm{d,n} = A_\mathrm{d,n}/(A_\mathrm{d,1}+A_\mathrm{d,2}+A_\mathrm{d,3})$. The fitting curves are superimposed to the data as solid lines in Fig.\ \ref{fig:2}(b) and the $\tau_\mathrm{d,n}$ values are displayed in the same figure (the complete set of the fitting parameters, including $w_\mathrm{d,n}$, can be found in \textcolor{purple}{Supporting Note 2}). The presence of different components (1, 2 and 3) indicates that different intermediate and intercommunicating levels are involved in the MX decay, possibly including dark exciton states \cite{Choi_IX_lifetime,time_resolved_novoselov}. In any case, $\tau_\mathrm{d,n}$ gets shorter for the higher energy ranges considered; this is particularly true for the 1.4 eV component, similar to recent results \cite{Choi_IX_lifetime,Wang_moire_mott,Tran:2019jl,time_resolved_novoselov}. This finding supports the hypothesis that the structured MX emission corresponds to a ladder of discrete states arising from the moir\'{e} potential \cite{Tran:2019jl,Choi_IX_lifetime,time_resolved_novoselov}. Indeed, higher-energy states may decay faster due to the  tendency of photo-generated carriers to occupy lower-lying states, with the ground state having the longest lifetime of several tens of ns, consistent with the spatially and k-space indirect characteristics of the MX transition.
We recall that in TMD MLs the intralayer exciton X is known to have much shorter recombination decay times, on the order of a few ps to a few of tens of ps \cite{decay_time_Robert,Blundo_prr}, in contrast with MX.
It is worth noting that the spectral range centred at 1.4 eV should be considered as a mixing of the highest energy level of the moir\'{e} potential and of the free IX component. The latter is indeed expected to have a shorter decay time due to its free-particle nature, as found in disordered semiconductors whenever localised and free excitons coexist \cite{vinattieri_InGaAsN}.

The different states of the moir\'{e} potential also present a different formation dynamics. Fig.\ \ref{fig:2}(c) shows the  time evolution of the MX signal up to 1 ns after the laser pulse excitation. In this case, the data are reproduced by
\begin{equation}
    I_\mathrm{rise}(t) = -A_\mathrm{r} \cdot \exp{\left( -\frac{t-t_0}{\tau_\mathrm{r}} \right)} + I_\mathrm{decay}(t)
    \label{eq:rise}
\end{equation}
where $\tau_\mathrm{r}$ is the luminescence rise time and $I_\mathrm{decay}$ represents the decay part of the data. By fitting the decay part first, the data in the (0-1) ns time interval can be reproduced by Eq. 2, with only $A_r$ and $\tau_r$ as fitting parameters.
The $\tau_\mathrm{r}$ values are displayed in panel (c) of Fig.\ \ref{fig:2} (the data corresponding to the high-energy range, shown in the right-most panel, could not be fitted reliably). The data indicate that the highest-energy excited state of the moir\'{e} potential (together with the likely presence of free IXs) is populated first ($<$100 ps), similar to what reported in Ref.\ \cite{time_resolved_novoselov}. Instead, the population of the lowest-energy state requires more time to reach its quasi-equilibrium occupancy because of the extra contribution from higher-energy levels in addition to the direct excitation.\\

\begin{figure*}[h!]
\includegraphics[width=\textwidth]{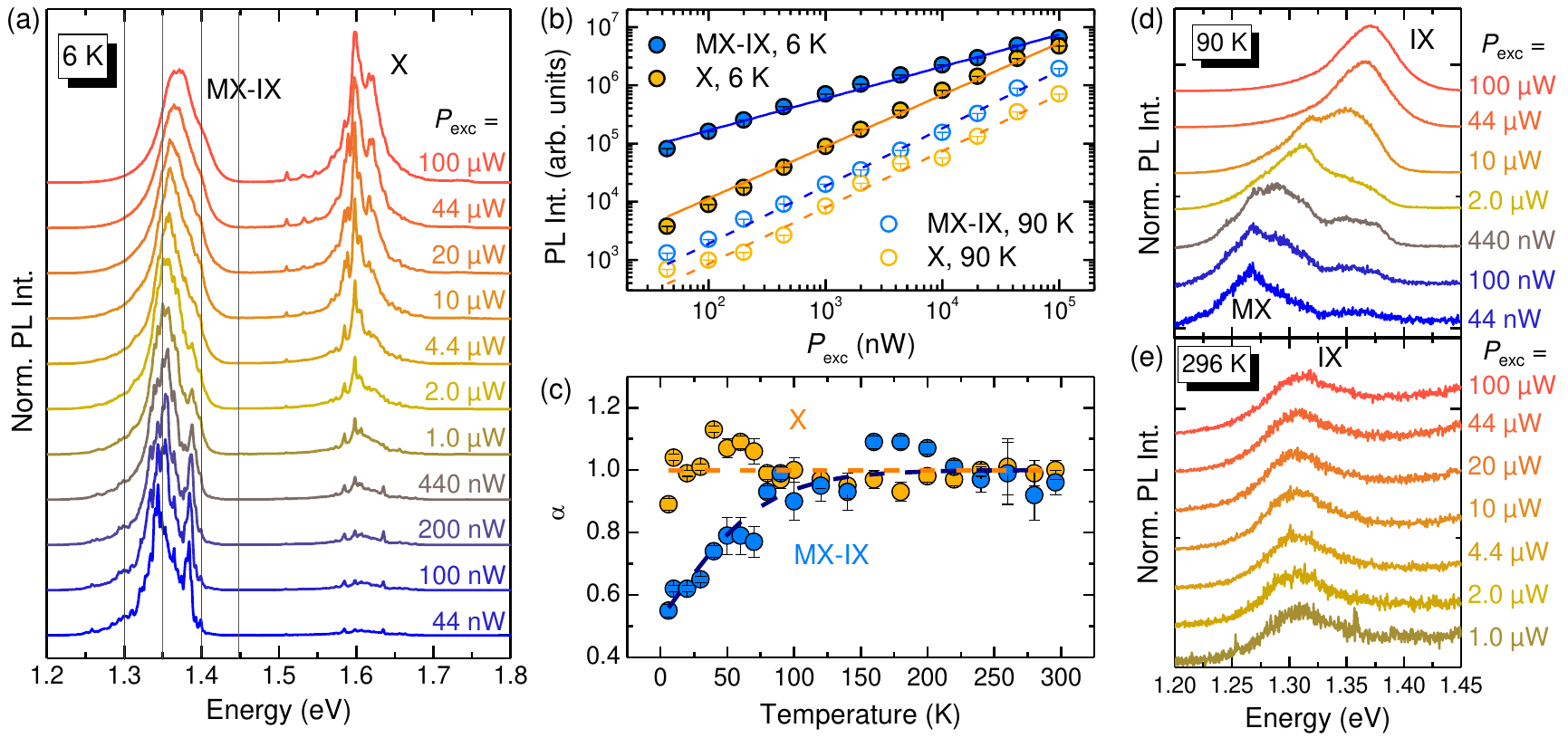}
\caption{\textbf{Photogenerated carrier density and temperature dependence of the exciton bands}. (a) $T$=6 K $\mu$-PL spectra of the studied WSe$_2$/MoSe$_2$ HS recorded for different laser excitation power values. MX indicates the moir\'{e} exciton band and X the intralayer exciton recombination in the MoSe$_2$ and WSe$_2$ layers (lower- and higher-energy side, respectively). (b) PL integrated intensity dependence on the laser power $P_\mathrm{exc}$ for MX (azure symbols) and X (dark yellow symbols) bands at $T$=6 K (full symbols) and $T$=90 K (open symbols). Solid and dashed lines are fits to the data with Eq.\ \ref{eq:IvsP} for $T$=6 K and 90 K, respectively. At $T$=6 K, the $\alpha$ coefficient values are $0.55 \pm 0.02$ and $0.89 \pm 0.02$ for MX and X, respectively. At $T$=90 K, the $\alpha$ coefficient values are $0.99 \pm 0.02$ and $0.97 \pm 0.03$ for MX and X, respectively. (c) Temperature variation of the $\alpha$ coefficient for the MX-IX and X bands. In the former case, a clear transition from a sublinear to a linear behaviour is found and ascribed to the transition from a moir\'{e} localisation regime to a free interlayer exciton one (hence the mixed label MX-IX). (d) $T$=90 K $\mu$-PL spectra for different laser excitation powers in the energy region where the MX and IX recombinations can be simultaneously observed. IX takes over MX upon increase of the photogenerated carrier density. (e) Same as (d) for $T$=296 K, where only the IX transition is observable.}
\label{fig:3}
\end{figure*}

\noindent
\textbf{Exciton recombination evolution with carrier density and temperature}\\
The X and MX recombination bands also exhibit  quite distinct spectral behaviours when the density of photogenerated excitons and the lattice temperature are increased. Fig.\ \ref{fig:3}(a) shows the cw $\mu$-PL spectra at $T$=6 K for $P_\mathrm{exc}$ ranging from 44 nW (\emph{i.e.} 5.6 W/cm$^2$) to 100 $\mu$W (\emph{i.e.} $1.3\cdot 10^{4}$ W/cm$^2$). The MX band broadens and its centroid blueshifts with increasing $P_\mathrm{exc}$, likely as a consequence of the dipole-dipole interaction between MXs \cite{Wang_moire_mott,brotons_prx,Seyler_moire_excitons,Nagler_PLvsT,Li_IX_PL_saturation}. As pointed out when describing Fig.\ \ref{fig:2}(a), the component centred at about 1.4 eV can be attributed to de-trapped IXs, as it will be confirmed next. Following Ref.\ \cite{Wang_moire_mott}, we determine that in the $P_\mathrm{exc}$=(0.044-100) $\mu$W range the density of photogenerated electron-hole pairs within the HS varies from $n_\mathrm{e-h}$=$1.1\cdot 10^{11}$ cm$^\mathrm{-2}$ to $2.3\cdot 10^{13}$ cm$^\mathrm{-2}$ (see \textcolor{purple}{Supporting Note 3}). We note that the highest $n_\mathrm{e-h}$ achieved by us is smaller than the value  necessary to observe an optically induced Mott transition from IXs to spatially separated electron and hole gases \cite{Wang_moire_mott}. Nevertheless, from the previously estimated period of the moir\'{e} potential $a_\mathrm{m}$= 40 nm, the corresponding density of moir\'{e} minima is equal to $7.2\cdot 10^{10}$ cm$^\mathrm{-2}$ and a sizeable exciton-exciton interaction is possible thus explaining the decrease in the emission decay time observed in Fig.\ \ref{fig:2}(b) as well as the MX band blushift with $P_\mathrm{exc}$  \cite{Wang_moire_mott,brotons_prx}. On the other hand, the X band, which, as we recall, comprises the MoSe$_2$ and WSe$_2$ intralayer excitons, does not change appreciably its centroid. It instead gains significant spectral weight compared to MX, which originates from recombination centres with finite spatial density. Fig.\ \ref{fig:3}(b) shows the dependence of the integrated intensity $I$ of the HS exciton (that we indicate as MX-IX to take into account also the contribution from IXs at high $P_\mathrm{exc}$, as discussed for Fig.\ \ref{fig:2}) and X bands as a function of $P_\mathrm{exc}$ for $T$=6 K and $T$=90 K. The data were fitted by:
\begin{equation}
    I = A\cdot{P_\mathrm{exc} ^{~\alpha}},
    \label{eq:IvsP}
\end{equation}
where $A$ is a scaling constant. At $T$=6 K, $\alpha$ is equal to $0.55 \pm 0.02$ for MX-IX and $0.89 \pm 0.02$ for X. The smaller $\alpha$ found for the MX-IX signal from the HS (as opposed to that from intralayer excitons X in the MLs) is compatible with the finite number of the moir\'{e} potential minima and with exciton-exciton interactions as a probable source for the signal loss. Instead, the nearly linear behaviour of the X emission intensity is consistent with the virtually unlimited number of intralayer excitons that can be photogenerated.
Interestingly, Fig.\ \ref{fig:3}(b) shows that the nearly linear dependence of the X band on $P_\mathrm{exc}$ is maintained also at $T$=90 K, while a major variation is found for the MX-IX band due to the increasingly higher spectral contribution of the free IXs at higher $T$. As a matter of fact, the $\alpha$ value of MX-IX becomes approximately equal to 
1 at 90 K.

%

 Puzzled by this finding, we investigated the dependencies on $P_\mathrm{exc}$ of the integrated area of the MX-IX and X bands at different temperatures. The full set of power-dependent data can be found in \textcolor{purple}{Supporting Note 4}. Fig.\ \ref{fig:3}(c) summarises the variation of the coefficient $\alpha$ with $T$, as obtained from Eq.\ \eqref{eq:IvsP}. For the X band, a nearly linear behaviour is observed at all temperatures. Instead, for the MX-IX band, $\alpha$ increases progressively from 0.55 to about 1 as $T$ is increased from 5 K to 120 K, with a linear behaviour observed at higher temperatures, up to room temperature. These results suggest a qualitative change in the nature of the exciton-related bands in the HS at about 120 K.
 
 Figs.\ \ref{fig:3}(d) and (e) display a series of spectra recorded at different $P_\mathrm{exc}$ for $T$=90 K and $T$=296 K, respectively. In the first case, the lineshape of the MX band changes significantly as the density of photoexcited carriers increases. Indeed, we notice a considerable spectral weight transfer from the structured band below 1.32 eV to the single component peaked at about 1.38 eV. We ascribe this change to the saturation of the moir\'{e}-localised excitons in favour of moir\'{e}-de-trapped IXs. This behaviour is not evident at the lowest $T$ values (see, \emph{e.g.}, Fig.\ \ref{fig:3}(a)), when the MXs are frozen in their potential minima. Eventually, for $T>200$ K, almost all MXs are ionised and only IXs are observed, as shown in Fig.\ \ref{fig:3}(e), clearly demonstrating the absence of a sizeable lineshape variation with $P_\mathrm{exc}$.\\ 

\begin{figure*}[h!]
\centering
\includegraphics[width=12 cm]{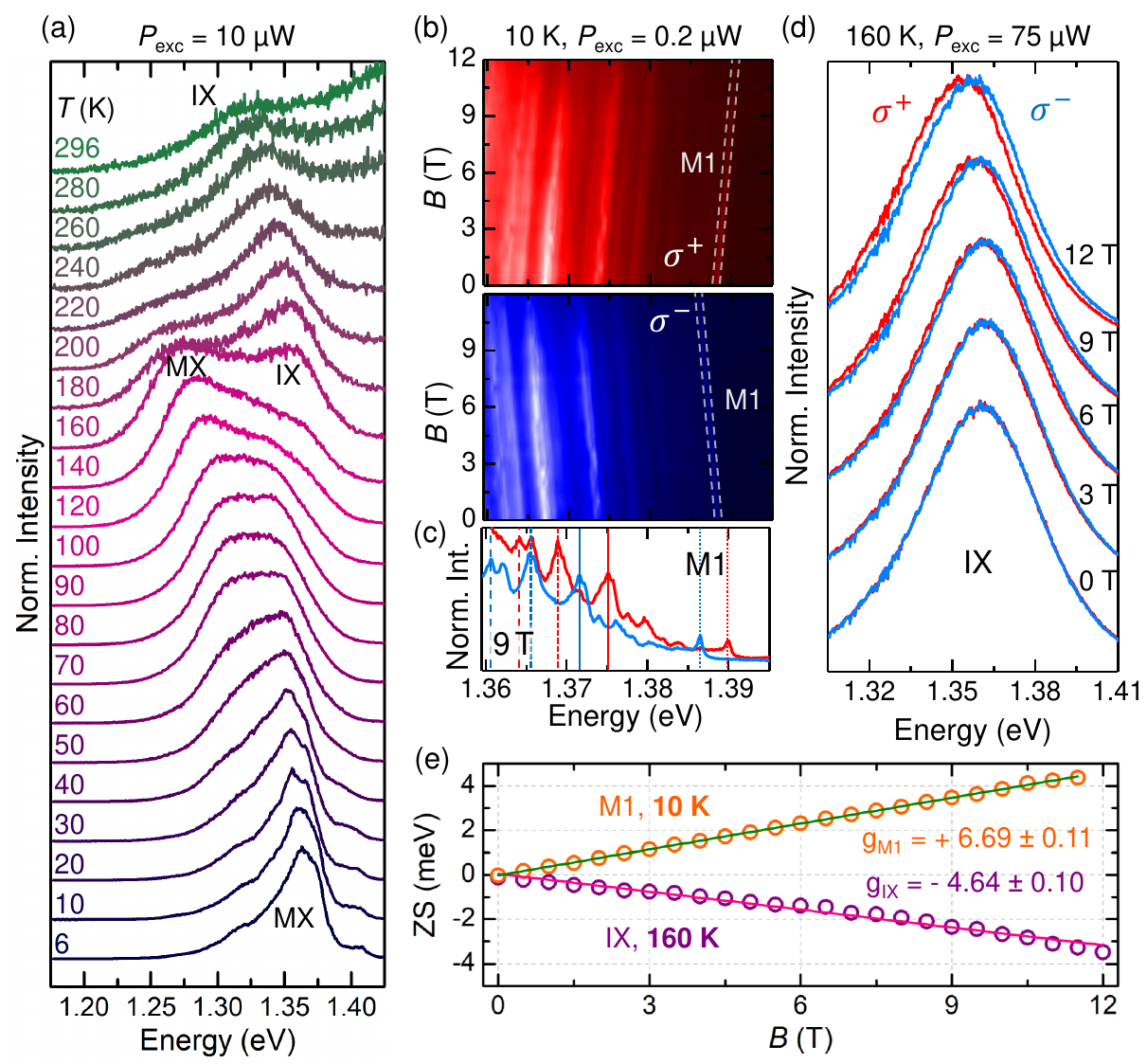}
\caption{\textbf{Exciton magnetic moment sign reversal}. (a) $\mu$-PL spectra of the studied WSe$_2$/MoSe$_2$ HS recorded for different temperatures and fixed $P_\mathrm{exc}$=10 $\mu$W focused via a 20$\times$ objective (NA=0.4). MX indicates the moir\'{e} exciton band and IX the free intralayer exciton recombination. Note the major spectral transfer from MXs to IXs for $T>120$ K. (b) Magneto-$\mu$-PL colour plots of the MX band at $T=10$ K and $P_\mathrm{exc}$=0.2 $\mu$W. The upper and lower panels correspond to $\sigma^\mathrm{+}$ (red) and $\sigma^\mathrm{-}$ (blue) circular polarisations, respectively. The positive and negative slopes of the $\sigma^\mathrm{+}$ and $\sigma^\mathrm{-}$ polarisations with the field indicate a positive gyromagnetic factor. M1 denotes one specific MX line, whose spectra at $B$=9 T are shown in panel (c) for opposite circular polarisations. (d) Magneto-$\mu$-PL spectra at $T=160$ K and $P_\mathrm{exc}$=75 $\mu$W of the free IX band for $\sigma^\mathrm{+}$ and $\sigma^\mathrm{-}$ polarisations. A negative ZS can be observed with the $\sigma^\mathrm{+}$ and $\sigma^\mathrm{-}$ spectra being at lower and higher energy, respectively. (e) ZS of the moir\'{e}-localised exciton M1 and of the free IX exciton \emph{vs} magnetic field resulting in the gyromagnetic factors displayed in the figure.}   
\label{fig:4}
\end{figure*}

\noindent
\textbf{Moir\'{e} exciton de-trapping, magnetic moment and dynamics with increasing lattice temperature}\\
The moir\'{e} exciton de-trapping is even more evident in Fig.\ \ref{fig:4}(a), which shows the $\mu$-PL spectra for different $T$ values and $P_\mathrm{exc}$=10 $\mu$W (corresponding to $n_\mathrm{e-h}$=$5.6\cdot 10^{12}$ cm$^\mathrm{-2}$); similar studies for higher and lower $P_\mathrm{exc}$ are shown in \textcolor{purple}{Supporting Note 5}. From $T$=6 K to $T$=120 K, the HS signal is dominated by the MX band, which undergoes a redistribution of the carrier population  between the different states of the moir\'{e} potential. Starting from $T$=120 K, a high-energy component due to IXs appears and becomes increasingly important relative to the MX band, until the latter vanishes at about 220 K. Finally, at room temperature only IXs are visible. At $T\approx160$ K, the two contributions coexist so that their energy difference can be estimated. The obtained value, equal to about 80 meV, fits well with the exciton barrier height of the moir\'{e} potential in R-type WSe$_2$/MoSe$_2$ HSs \cite{LI_IX_transport,Tran:2019jl}, where only the exciton singlet state is optically permitted. In contrast, the exciton ground state in H-type HSs is in a triplet configuration, with the singlet state having an energy 25 meV higher \cite{Seyler_moire_excitons,brotons_prx,Zhang_singlet_triplet}. We exclude that the two transitions coexisting at intermediate $T$ are ascribable to $K_\mathrm{CB}$-$K_\mathrm{VB}$ (CB and VB stand for conduction and valence band, respectively) and $\Lambda_\mathrm{CB}$-$K_\mathrm{VB}$ IX transitions \cite{Barre_IX_absorption}, which differ by 55 meV \cite{Barre_IX_absorption}. It is worth mentioning that different results were reported. In Ref.\ \cite{MX_vs_T_Schaibley}, the MX de-trapping was observed by monitoring the PL intensity and lifetime of WSe$_2$/MoSe$_2$ HSs with a transition temperature $<50$ K that is in contrast with our results. On the other hand, exciton diffusivity measurements \cite{LI_IX_transport} showed the absence of MX de-trapping in a WSe$_2$/MoSe$_2$ HS with nearly perfect lattice alignment ($ \theta= 0.15^\circ$), while a clear de-trapping was visible for $ \theta>2^\circ$  \cite{LI_IX_transport}.

In any case, the observed temperature-induced change in the nature of the exciton in the HS should be reflected in the electronic properties of the levels involved in the exciton recombination. In this respect, the exciton magnetic moment and the associated gyromagnetic factor $g_\mathrm{exc}$ ---embedding the spin, orbital and valley properties of the bands---  turned out to be an extremely sensitive parameter of the electronic structure of nanostructures \cite{magnetoPL_NW} and of 2D crystals \cite{MagneticField_WSe2_stier,Molas2019PRL,Blundo_magnetoPL,Kasia_WSSe} and their HSs \cite{Wozniak_g_factor_HS,Ciarrocchi_g_factor_HS,Seyler_moire_excitons,brotons_prx,Joe_g_factor_HS,Li_IX_PL_saturation,Li_localized_IX,Liu_moire_lattice_period,Nagler_magnetoPL,Holler_g_factor_HS}. In WSe$_2$/MoSe$_2$ HSs, the lowest-energy exciton state is in a spin-singlet configuration for R-type HSs and in a spin-triplet configuration for H-type HSs \cite{Yu_theory_HS}. The spin-singlet and spin-triplet excitons feature a $g_\mathrm{exc}$ value with a positive ($\approx+7$) and a negative ($\approx-15$) sign, respectively, the exact value depending on the specific sample \cite{Ciarrocchi_g_factor_HS,Seyler_moire_excitons,Joe_g_factor_HS,brotons_prx,Holler_g_factor_HS,Nagler_magnetoPL,Wang_singlet_triplet,Wozniak_g_factor_HS}. Our HS is R-type, as discussed before, and therefore we expect a positive $g_\mathrm{exc}$ value. Fig.\ \ref{fig:4}(b) shows the magnetic field, $B$, dependent $\mu$-PL spectra in the HS exciton region at $T=10$ K and $P_\mathrm{exc}$=200 nW (corresponding to $n_\mathrm{e-h}$=$4.0\cdot 10^{11}$ cm$^\mathrm{-2}$, see Supporting Note 3). For each field, the opposite circular polarisations  $\sigma^\mathrm{\pm}$ were recorded simultaneously on two different regions of the CCD detector, see Methods, and are shown separately in Fig.\ \ref{fig:4}(b). Different MXs are observed exhibiting a \emph{positive} Zeeman splitting (ZS), defined as
\begin{equation}
\mathrm{ZS}(B) = E^{\sigma^+} - E^{\sigma^-} = g_\mathrm{exc} \cdot \mu_\mathrm{B} B.
\label{eq:ZS}
\end{equation}
$E^{\mathrm{\sigma^{\pm}}}$ are the peak energies of components with opposite helicity $\sigma^+$ and $\sigma^-$, and $\mu_\mathrm{B}$ is the Bohr magneton. The positive (negative) slope with $B$ of the $\sigma^+$ ($\sigma^-$) component of the lines displayed in Fig.\ \ref{fig:4}(b) indicates that $g_\mathrm{exc}>0$ for individual MXs (see also Fig.\ \ref{fig:4}(c), where the $\sigma^+$ red component is at higher energy than the $\sigma^-$ blue one). Then, magneto-$\mu$-PL measurements were performed also at $T=160$ K ($P_\mathrm{exc}$=75 $\mu$W), where the HS exciton band is instead dominated  by free IXs. Fig.\ \ref{fig:4}(d) shows the $\sigma^+$ and $\sigma^-$ components of the IX spectra at different magnetic fields. Remarkably, the two components exhibit a \emph{negative} ZS, \emph{i.e.} opposite to that found at $T=10$ K for the MX lines (in this case, the $\sigma^+$ red component is at lower energy than the $\sigma^-$ blue one). Fig.\ \ref{fig:4}(e) shows the ZS field dependence for one MX line at 10 K (circles) and for the IX band at 160 K (squares), both fitted (lines) by  Eq.\ \eqref{eq:ZS}. The resulting $g_\mathrm{exc}$ for the (trapped) MXs and (free) IXs are $g_{exc,\mathrm{MX}} = +6.69 \pm 0.11$ and $g_{\mathrm{exc},\mathrm{IX}} = -4.64 \pm 0.10$, respectively. The former is in close agreement with previous experimental \cite{Seyler_moire_excitons,Joe_g_factor_HS,Li_localized_IX,Ciarrocchi_g_factor_HS} and theoretical \cite{Wozniak_g_factor_HS} results found for MXs in R-type WSe$_2$/MoSe$_2$ HSs. As for the results at 160 K, to our knowledge there are no previous ZS measurements at high temperatures. We found a similar $g_{\mathrm{exc},\mathrm{IX}}$ value also at $T$=210 K and 100 K, as described in \textcolor{purple}{Supporting Note 6}. The origin of the sign reversal of $g_{\mathrm{exc},\mathrm{IX}}$ must then be ascribed to the avoided effect of the moir\'{e} potential caused by the temperature-induced  de-trapping of the MXs. As a matter of fact, we can estimate $g_{\mathrm{exc},\mathrm{IX}}$ considering the separate contribution of electrons and holes to the IX gyromagnetic factor,
as usually done for excitons in semiconductors. Following an analogous procedure to that employed in Ref.\ \cite{Blundo_magnetoPL} for strained WS$_2$ MLs, for this HS we use
\begin{equation}
g_{\mathrm{exc},\mathrm{IX}} = 2\left[ L_\mathrm{CB}(\mathrm{MoSe}_2) - L_\mathrm{VB}(\mathrm{WSe}_2) \right],
\label{eq:gexc}
\end{equation}
where the first and second terms are the expectation values of the orbital angular momentum of the  MoSe$_2$ CB and WSe$_2$ VB, respectively (the spin contribution cancels out because the band extrema involved in the free IX transition have the same spin for R-type HSs). As reported in Ref.\ \cite{Wozniak_g_factor_HS}, $L_\mathrm{CB}(\mathrm{MoSe}_2)$=1.78 and $L_\mathrm{VB}(\mathrm{WSe}_2)=4.00$ and from Eq.\ \eqref{eq:gexc} we obtain $g_{\mathrm{exc},\mathrm{IX}}=-4.44$ in very good agreement with the value we found experimentally for the free IX transition shown in Fig.\ \ref{fig:4}(d). Interestingly, the suppression of the moir\'{e} potential in a R-type WSe$_2$/MoSe$_2$ HS caused by inserting a h-BN layer between the constituent MLs leads to magneto-PL results very similar to ours \cite{localized_IX_vs_defects}. Indeed, in Ref.\ \cite{localized_IX_vs_defects} the spatial decoupling between the HS single layers determines a sign reversal and a decrease of the $g_\mathrm{exc}$ modulus analogous to that found here by increasing the lattice temperature. Likewise, a sign reversal of $g_\mathrm{exc}$ can be observed by increasing $P_\mathrm{exc}$, and hence the density of electron-hole pairs. Under these circumstances, $g_\mathrm{exc}$ becomes negative when the entire ensemble of (interacting) IXs is considered, as shown in \textcolor{purple}{Supporting Note 7}. Although this finding requires further investigations, we ascribe this effect to an effective screening of the moir\'{e} potential caused by photogenerated carriers \cite{brotons_prx}.

On the one hand, these observations confirm the profound effect that the moir\'{e} potential exerts on the exciton physics of the HS up to high temperatures. On the other hand, magnetic fields represent thus a quite valuable tool to determine the localised or delocalised status of charge carriers in 2D HSs, which can be important for the understanding of fundamental effects, such as the formation of highly correlated electronic phases \cite{Zang_BE_IXcondensation,Wang_moire_mott,Mott_moire_Regan,Mott_moire_Tang}.

\begin{figure*}[h!]
\centering
\includegraphics[width=14.0cm]{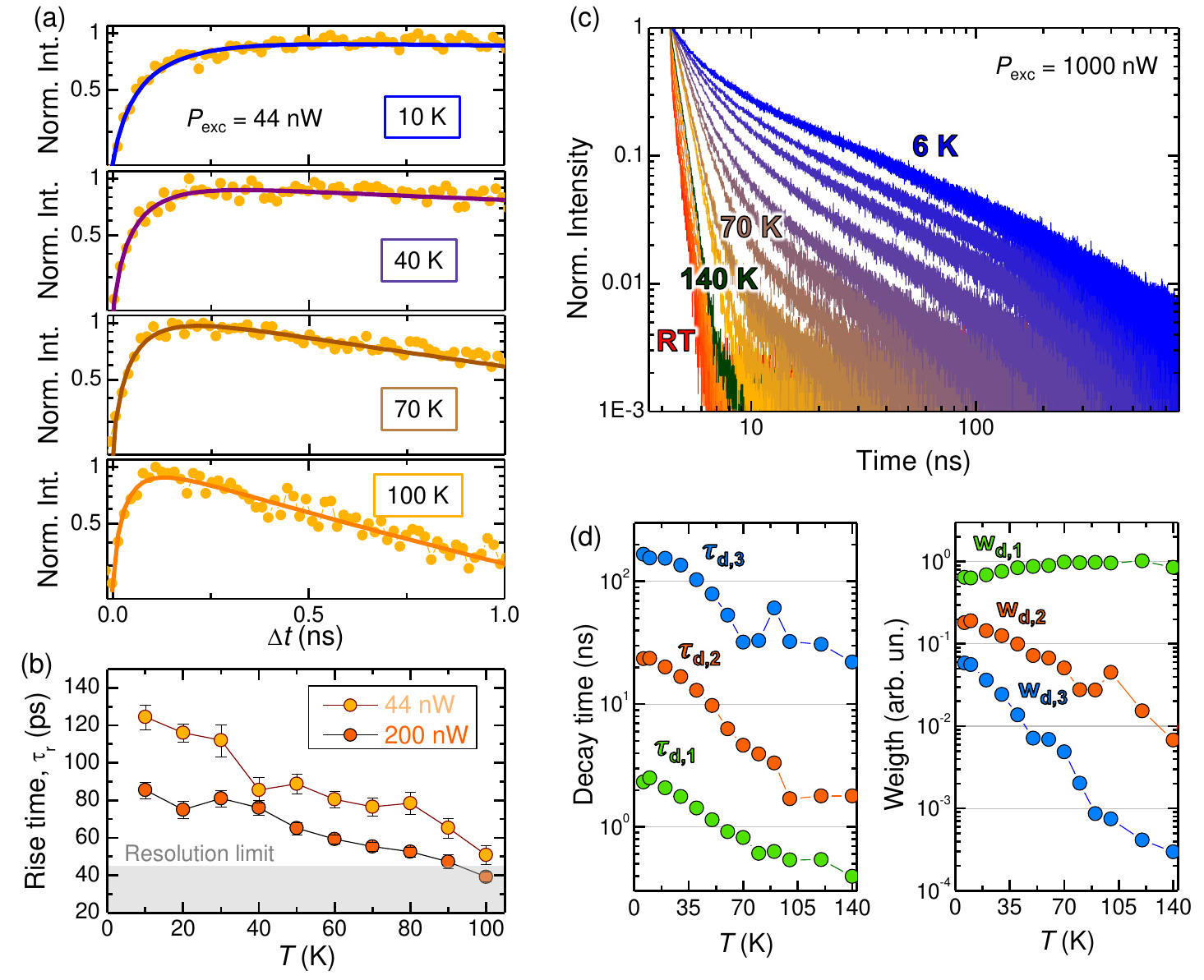}
\caption{\textbf{Rise and decay times with increasing temperature}. (a) Time-evolution of the $\mu$-PL signal of the investigated WSe$_2$/MoSe$_2$ HS recorded at different temperatures (and fixed laser excitation power $P_\mathrm{exc}$) in the $\Delta$t=0-1.0 ns interval from the laser pulse. The detection energy was set at the MX-IX band (see Fig.\ \ref{fig:3}). The laser power was 44 nW. The solid lines are fits to the data by Eq.\ \ref{eq:rise}. (b) Rise time $\tau_\mathrm{r}$ values obtained by fitting the experimental data for different temperatures and two $P_\mathrm{exc}$ values. The setup time resolution is shown by the grey area. Notice that once the data get close to the resolution limit, the estimated rise time is affected by the system response and thus only qualitatively indicative. (c) Time-evolution of the $\mu$-PL signal of the MX-IX band (see Fig.\ \ref{fig:3}) recorded in the $\Delta$t=0-800 ns interval from the laser pulse. The data were recorded at different temperatures as indicated in the figure and fixed $P_\mathrm{exc}$. (b) Decay times $\tau_\mathrm{d,n}$ values used to reproduce the data via Eq.\ \ref{eq:decay}. The same for the spectral weights $w_\mathrm{n}$ of the different time components (Eq.\ \ref{eq:decay}).}
\label{fig:5}
\end{figure*}

Related to the previous discussion is the change in the MX formation and recombination dynamics when the de-trapping process starts to occur with increasing $T$. Fig.\ \ref{fig:5}(a) displays the time evolution of the $\mu$-PL exciton signal in the HS within 1 ns after the laser excitation, which corresponds predominantly to the exciton formation. Different temperatures were considered with $P_\mathrm{exc}$=44 nW ($n_\mathrm{e-h}$=$1.1\cdot 10^{11}$ cm$^\mathrm{-2}$).
It is clear that the MX formation dynamics becomes faster with increasing $T$ (for $T\leq$100 K and $P_\mathrm{exc}$=44 nW, MXs dominate). The experimental data were fitted by Eq.\ \eqref{eq:rise} and the $T$ dependence of $\tau_\mathrm{r}$ is displayed in panel (b) for two different photogenerated carrier densities. At $T$=100 K, $\tau_\mathrm{r}$ approaches the temporal resolution limit (notice that once the data get close to the resolution limit, the estimated rise time is affected by the system response and thus only qualitatively indicative). The higher temperatures and the ensuing MX ionisation process result indeed in a decreased contribution of the moir\'{e} localisation step and thus in a reduction of the time required to build up the exciton population contributing to the MX/IX band. This process is more evident with increasing $P_\mathrm{exc}$, as can be noted in Fig.\ \ref{fig:5}(b). As a matter of fact, a larger photogenerated carrier density tends to saturate the MX states shifting the spectral centroid of the MX-IX band towards the faster-forming IX levels. 

The luminescence decay is also highly influenced by temperature variations. Fig.\ \ref{fig:5}(c) shows the MX-IX band decay curves at $P_\mathrm{exc}$=1 $\mu$W ($n_\mathrm{e-h}$=$1.2\cdot 10^{12}$ cm$^\mathrm{-2}$) and different temperatures. The curves can be reproduced using Eq.\ \eqref{eq:decay} and the values of the fitting parameters  ($\tau_\mathrm{d,n}$ and $w_\mathrm{d,n}$) are displayed in Fig.\ \ref{fig:5}(d). The three values of the decay time $\tau_\mathrm{d,n}$ decrease monotonically, with the shorter one ($\tau_\mathrm{d,1}$) reaching the resolution limit (0.23 ns) at $T$=140 K, and the weights of the slower components ($w_\mathrm{d,2}$ and $w_\mathrm{d,3}$) becoming less important. The latter are particularly relevant at low $T$, where decay time values of about 200 ns are observed, consistent with the space-indirect nature of the MXs. The, yet small, k-space mismatch associated with the twist angle may also contribute to the lengthening of the luminescence decay time \cite{Choi_IX_lifetime}. The marked decrease of $\tau_\mathrm{d,n}$ with $T$ can be explained by two simultaneous mechanisms. First, non-radiative recombination channels are activated at higher temperatures, greatly shortening the luminescence decay time. Second, delocalised states are expected to have a larger recombination probability, because they are more likely to interact with other oppositely charged free carriers, or with lattice defects acting as non-radiative channels. 

\section*{Conclusions}

We investigated the process of temperature-induced exciton de-trapping from moir\'{e} minima in a WSe$_2$/MoSe$_2$ HS.
We observed that at $T\approx120$ K moir\'{e} excitons turn into free interlayer excitons with relevant consequences for quantum technology applications \cite{Baek:2020cf} and for the observation of many-body phenomena, such as exciton condensation \cite{Zang_BE_IXcondensation} or Mott transition \cite{Wang_moire_mott,Mott_moire_Regan,Mott_moire_Tang}. The temperature-induced transition from a moir\'{e}-confined to a free IX regime manifests in a sizeable variation of the power law governing the exciton signal growth with photogenerated carrier density. The exciton magnetic moment too undergoes major variations with increasing \emph{T}. Indeed, the interlayer exciton $g$-factor exhibits a remarkable reversal of its sign and decrease of its modulus (going from $\sim +7$ to $\sim -5$) concomitantly with the de-trapping of the moir\'{e}-confined excitons for $T \gtrsim 120$ K. This may also have relevant consequences for valleytronic applications of TMD HSs. Within the same $T$ interval, we also consistently found that the formation time of MXs is strongly reduced as a consequence of the cross-over from a localised to a free-like regime. This indicates that the exciton capture in the moir\'{e} potential requires an intermediate step that lengthens the luminescence rise time. Also, the decay time of the MX/IX states is greatly reduced by increasing $T$ due to the increased recombination probability of freely moving excitons as well as to exciton-exciton interactions and to thermally activated non radiative recombination channels.
Our findings shed new light on the truly confined nature of the exciton states in a moir\'{e} superlattice with increasing temperature thus setting the conditions for the observation and stability of highly correlated phases at elevated temperatures in moir\'{e} superlattices.

\vspace{1.0cm}

\section*{Methods}

\noindent
\textbf{Sample fabrication}\\
The heterostructure (HS) was fabricated by the standard dry transfer technique. TMD flakes were mechanically exfoliated by the scotch tape method and deposited on PDMS. MoSe$_2$ and WSe$_2$ monolayers on the PDMS were identified and deposited on a SiO$_2$/Si substrate.
The MoSe$_2$ was deposited first, and the WSe$_2$ was deposited atop of it. The sample was annealed in high vacuum at 150 $^\circ$C for some hours. 
hBN flakes were then exfoliated with the same approach and a thin hBN flake was identified on the PDMS. The flake was then deposited in such a way to cap the HS completely.
The sample was re-annealed under the same conditions.\\


\noindent
\textbf{Continuous-wave $\mu$-PL measurements}\\
For $\mu$-PL measurements, the excitation laser was provided by a single frequency Nd:YVO$_4$ lasers (DPSS series by Lasos) emitting at \SI{532}{\nano\metre}. The luminescence signal was spectrally dispersed by a \SI{20.3}{\centi\metre} focal length Isoplane 160 monochromator (Princeton Instruments) equipped with a 150 grooves/mm and a 300 grooves/mm grating and detected by a back-illuminated N$_2$-cooled Si CCD camera (100BRX by Princeton Instruments). The laser light was filtered out by a very sharp long-pass Razor edge filter (Semrock). A 100$\times$ long-working-distance Zeiss objective with NA = 0.75 was employed to excite and collect the light, in a backscattering configuration using a confocal setup.\\
For high resolution measurements aimed at\\ 
higlighting the moir\'{e} energy levels (Fig.\ 1(c)), a 75 cm focal length Acton monochromator was used.\\

\noindent
\textbf{Time-resolved $\mu$-PL measurements}\\
For tr $\mu$-PL measurements, the sample was excited with a ps supercontinuum laser (NKT Photonics) tuned at 530 nm, with a full width at half maximum of about 10 nm and 50 ps pulses at 1.2 MHz repetition rate. The sample was excited in the same experimental configuration used for continuous wave measurements. The signal was then collected in a backscattering configuration and the desired spectral region was selected by using longpass and shortpass filters. The signal was thus focused by means of a lens on an avalanche photodetector from MPD with temporal resolution of 30 ps. \\

\noindent
\textbf{$\mu$-PL excitation measurements}\\
For $\mu$-PL excitation ($\mu$-PLE), we employed the same ps supercontinuum laser used for tr $\mu$-PL. The laser wavelength was automatically changed by an acousto-optic tunable filter and employing a series of shortpass and longpass filters to remove spurious signals from the laser. The detection wavelength was selected using the same monochromator and detector employed for cw $\mu$-PL measurements.\\

\noindent
\textbf{Magneto-$\mu$-PL measurements}\\ Magneto-$\mu$-PL measurements were performed at variable temperature in a superconducting magnet reaching up to 12 T. x-y-z piezoelectric stages were used to excite the sample and collect the signal from the desired point of the sample. A 515-nm-laser and a 100$\times$ microscope objective with NA=0.82 were used. The same objective was used to collect the luminescence. The circular polarisation of the luminescence was analysed using a quarter-wave plate (that maps circular polarisations of opposite helicity into opposite linear polarisations) and a Wollaston prism steering the components of opposite linear polarisation (and thus of opposite helicity) to different lines of the liquid-nitrogen-cooled Si-CCD we employed (100BRX by Princeton Instruments). In this manner, the $\sigma^+$ and $\sigma^-$ components could be measured simultaneously. A monochromator with 0.75 m focal length (Princeton Instruments) and a 600 grooves/mm grating was used to disperse the PL signal. 

The field was directed perpendicular to the sample surface ($i.e.$, parallel to the emitted photon wavevector, Faraday configuration).

\section*{Acknowledgments}
The authors thank Paulo E.\ Faria Jr and Tomasz Wo\'{z}niak for useful discussions.
We acknowledge support by the European Union's Horizon 2020 research and innovation programme through the ISABEL project (No. 871106). This project was funded within the QuantERA II Programme that has received funding from the European Union’s Horizon 2020 research and innovation programme under Grant Agreement No 101017733, and with funding organisations Ministero dell'Universit\'{a} e della Ricerca (MUR) and by Consiglio Nazionale delle Ricerche (CNR). E.B. acknowledges support from La Sapienza through the grants Avvio alla Ricerca 2021 (grant no. AR12117A8A090764) and Avvio alla Ricerca 2022 (grant no.\ AR2221816B672C03).
The authors acknowledge support from the National Science Centre, Poland, through Grants No. 2018/31/B/ST3/02111 (K.O.‐P. and M.R.M.) and No. 2017/27/B/ST3/00205 (A.B.).
K.W. and T.T. acknowledge support from the JSPS KAKENHI (Grant Numbers 19H05790 and 20H00354).

\section*{Author contributions}
E.B. and A.P. conceived and supervised the research. E.B. and M.C. fabricated the heterostructures. E.B., F.T., S.C., M.C. and G.C. performed the optical measurements and analysed the data. E.B., A.P., K.O.P., L.K., and M.R.M. performed the magneto-optical measurements, with the support of A.B., and E.B. analysed the data. A.M. provided support for optical measurements. M.F. provided support for the time-resolved measurements. G.P. contributed to the sample characterisation. T.T. and K.W. grew the hBN samples. E.B. and A.P. wrote the manuscript. The results and the manuscript were approved by all the coauthors.\\

\onecolumn


\begin{titlepage}

\begin{center}
	{\Large SUPPORTING INFORMATION for\\\mbox{}\\
		\textbf{
		Localisation-to-delocalisation transition of moir\'{e} excitons in WSe$_2$/MoSe$_2$ heterostructures
		}}
\end{center}

\vspace{1 mm}

\begin{center}
	{\large Elena Blundo,$^{1,*}$  Federico Tuzi,$^{1}$ Salvatore Cianci,$^{1}$ Marzia Cuccu,$^{1}$ Katarzyna Olkowska-Pucko,$^{2}$ \L{}ucja Kipczak,$^{2}$ Giorgio Contestabile,$^{1}$ Antonio Miriametro,$^{1}$ Marco Felici,$^{1}$ Giorgio Pettinari,$^{3}$ Takashi Taniguchi,$^{4}$ Kenji Watanabe,$^{5}$ Adam Babi\'{n}ski,$^{2}$ Maciej R. Molas,$^{2}$ and Antonio Polimeni$^{1,*}$}
\end{center}

\begin{center}
	\textit{\mbox{}$^1$ Physics Department, Sapienza University of Rome, 00185, Roma, Italy.\\
 \mbox{}$^2$ Institute of Experimental Physics, Faculty of Physics, University of Warsaw, Pasteura 5, 02-093 Warsaw, Poland\\
	\mbox{}$^3$ Institute for Photonics and Nanotechnologies (CNR-IFN), National Research Council, 00133, Rome, Italy\\
	\mbox{}$^4$ International Center for Materials Nanoarchitectonics, National Institute for Materials Science, 1-1 Namiki, Tsukuba 305-0044, Japan.\\
	\mbox{}$^5$ Research Center for Functional Materials, National Institute for Materials Science, 1-1 Namiki, Tsukuba 305-0044, Japan\\
	}
\end{center}

\mbox{}$^*$ Corresponding authors: antonio.polimeni@uniroma1.it, elena.blundo@uniroma1.it

\vspace{5 mm}

\fancyhead[LE,RO]{\tiny\thepage}
\renewcommand{\thesubfigure}{\alph{subfigure})}

\tableofcontents

\end{titlepage}

\newpage

\renewcommand{\theequation}{\arabic{section}.\arabic{equation}}
\renewcommand{\thetable}{\arabic{section}.\arabic{table}}
\renewcommand{\thefigure}{\arabic{section}.\arabic{figure}}
\renewcommand{\thesubsection}{A.\arabic{subsection}}
\setcounter{subsection}{0}
\renewcommand{\thesubfigure}{}
\sectionmark{} 


	\section*{Supporting Note 1. Moir\'{e} period and stacking angle}
	\addcontentsline{toc}{section}{Supporting Note 1. Moir\'{e} period and stacking angle}
	\setcounter{section}{1}
        \sectionmark{}
        \setcounter{equation}{0}

    The Hamiltonian for excitons confined in a moir\'{e} potential can be described as \cite{Wu_theory_moire_potential,Tran:2019jl}:
    \begin{equation}
        H = \hbar \Omega_0 + \frac{\hbar^2 k^2}{2 M} + \Delta(\mathbf{r}),
    \end{equation}
    where the first term is an energy constant, the second term is the center of mass kinetic energy, $M$ is the exciton mass and $\Delta(\mathbf{r})$ is the moir\'{e} potential energy.
    For a MoSe$_2$/WSe$_2$ HS, $M \approx 0.84 m_\mathrm{e}$, where $m_\mathrm{e}$ is the electron bare mass \cite{Tran:2019jl}.
    Near its minima, the moir\'{e} potential $\Delta(\mathbf{r})$ can be approximated as parabolic: 
    $\Delta(\mathbf{r}) = \beta (\mathbf{r} / a_M )^2 /2$, where $a_M$ is the moir\'{e} potential period and $\beta$ is a constant independent of $a_M$. Excitons confined in this parabolic potential have quantised energy levels:
    \begin{equation}
        E_m = \sqrt{ \frac{\beta \hbar^2}{M a_M^2}} \cdot (n_x + n_y + 1),
    \end{equation}
    where $n_{x,y}$ are non-negative integers \cite{Wu_theory_moire_potential}.
    The spacing between subsequent levels is thus:
    \begin{equation}
        S = \sqrt{ \frac{\beta \hbar^2}{M a_M^2}}.
    \end{equation}
    Following ref.\ \cite{Tran:2019jl}, $\beta = 2.84$ eV.
    
    Given that our moir\'{e} exciton levels are spaced by $(12.8 \pm 1.3)$ meV (see Fig.\ 1(c) of the main text), we estimate a period of
    \begin{equation}
        a_M = \frac{1}{S} \cdot \sqrt{ \frac{\beta \hbar^2}{M}} \approx 40~ \mathrm{nm}
    \end{equation}
    for our HS.
    
    According to the calculations by Liu \emph{et al.} \cite{Liu_moire_lattice_period}, such a period corresponds to a stacking angle of 0.5$^\circ$.

    \clearpage

	\newpage
	\subsection*{Supporting Note 2. Time-resolved $\mu$-PL data}
	\addcontentsline{toc}{section}{Supporting Note 2. Time-resolved $\mu$-PL data}
	\setcounter{section}{2}
        \sectionmark{}
	\setcounter{figure}{0}
	\setcounter{equation}{0}

	Here we display the fitting weights and decay times obtained by fitting the data in Fig.\ 2(b) of the main text.

 \begin{table}[h!]
\caption{Fitting parameter values obtained by fitting the data in Fig.\ 2(b) by Eq. (1) of the main text. $\tau_\mathrm{d,n}$ is the decay time relative to the $n$-th component, whose weight is given by $w_\mathrm{d,n}$.}
\begin{center}
{\renewcommand{\arraystretch}{1.2}
\begin{tabular*}{1.0\columnwidth}{@{\extracolsep{\fill}}ccccccc@{}}
\hline
\hline
Energy Range & $\tau_{\mathrm{d,1}}\;(\mathrm{ns})$ & $w_{\mathrm{d,1}}$ (\%) &  $\tau_{\mathrm{d,2}}\;(\mathrm{ns})$ & $w_{\mathrm{d,2}}$ (\%) & $\tau_{\mathrm{d,3}}\;(\mathrm{ns})$ & $w_{\mathrm{d,3}}$ (\%) \tabularnewline
\hline
Low (pink) & $2.27 \pm 0.09$ & $60.9 \pm 1.4$ & $14.0 \pm 0.6$ & $34.6 \pm 0.8$ & $51 \pm 4$ & $4.4 \pm 0.7$ \tabularnewline
Medium (cyan) & $1.59 \pm 0.01$ & $78.1 \pm 0.4$ & $15.2 \pm 0.2$ & $15.8 \pm 0.1$ & $77.0 \pm 0.3$ & $6.0 \pm 0.1$ \tabularnewline
High (yellow) & $<0.23$ & $86.0 \pm 0.4$ & $1.03 \pm 0.01$ & $13.0 \pm 0.3$ & $13.5 \pm 0.1$ & $1.0 \pm 0.1$ \tabularnewline
\hline
\end{tabular*}}
\end{center}
\label{tab:decay_3ranges_parameters}
\end{table}

 \clearpage

	\newpage
	\subsection*{Supporting Note 3. Estimation of photogenerated carrier density}
	\addcontentsline{toc}{section}{Supporting Note 3. Estimation of photogenerated carrier density}
	\setcounter{section}{3}
        \sectionmark{}
	\setcounter{figure}{0}
	\setcounter{equation}{0}
        \setcounter{table}{0}

	Figure 3 in the main text shows the cw $\mu$-photoluminescence (PL) spectra at $T$=6 K varying the laser power $P_\mathrm{exc}$ between 44 nW and 100 $\mu$W. To each $P_\mathrm{exc}$ value we associate a specific density of electron-hole pairs  $n_\mathrm{e-h}$ photogenerated within the HS. This was done following Ref. \cite{Wang_moire_mott}. Below we summarise the procedure followed.
	
	Under continuous wave (cw) excitation, the generation rate of photogenerated carriers is given by
	\begin{equation}
        G(n_\mathrm{e-h}) = \frac{P_\mathrm{exc} \cdot \sigma(n_\mathrm{e-h})}{S \cdot h\nu},
        \label{G}
    \end{equation}
		where $\sigma$($n_\mathrm{e-h}$) is the dependence of the absorbance of the MoSe$_2$/WSe$_2$ HS on the photogenerated carrier density,  $S=\pi r^{2}$ is the laser spot area with $r$=500 nm, and h$\nu$=2.33 eV is the exciting photon energy. In addition, in a stationary (\emph{i.e.} continuous wave) regime we have
		\begin{equation}
       \frac{\mathrm{d}n_\mathrm{e-h}}{\mathrm{d}t} = G(n_\mathrm{e-h}) - \frac{n_\mathrm{e-h}}{\tau(n_\mathrm{e-h})}=0,
        \label{n_vs_t}
    \end{equation}
		where $\tau$($n_\mathrm{e-h}$) is the dependence of the exciton decay time of the MoSe$_2$/WSe$_2$ HS on the photogenerated carrier density. To solve this equation we need to derive $\sigma$($n_\mathrm{e-h}$) and $\tau$($n_\mathrm{e-h}$).
		
\begin{figure}[!ht]
		\centering
		\includegraphics[width=9cm]{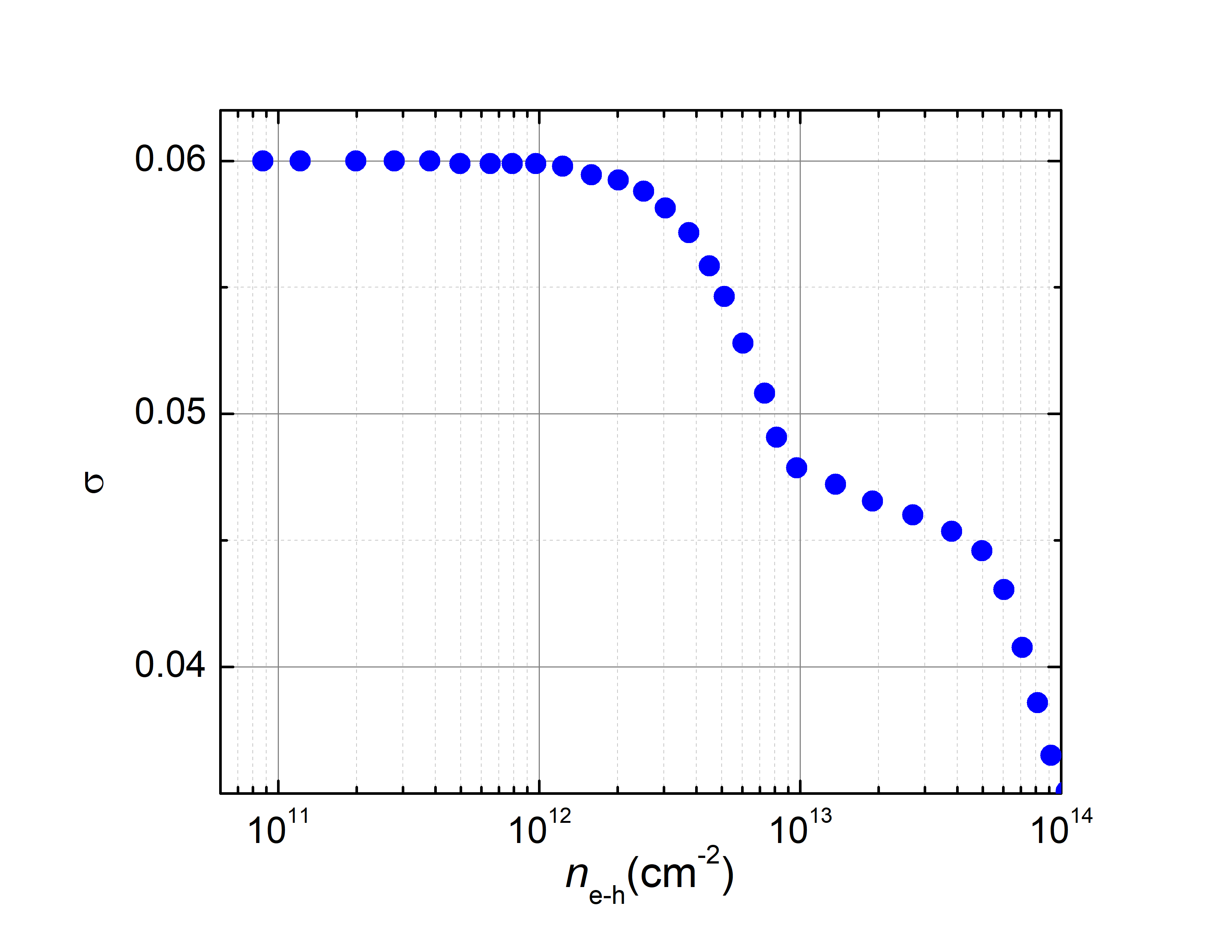}
		\caption{Optical absorbance of a MoSe$_2$/WSe$_2$ HS as a function of the density of photogenerated carriers for photon energy equal to 2.33 eV. The data were taken from Fig. 4 of Ref. \cite{Wang_moire_mott}.}
		\label{fig:absorbance}
	\end{figure}

		$\sigma$($n_\mathrm{e-h}$) was previously reported in Ref. \cite{Wang_moire_mott} and it is reproduced in Fig. \ref{fig:absorbance}.
	
\begin{figure}[!ht]
		\centering
		\includegraphics[width=\textwidth]{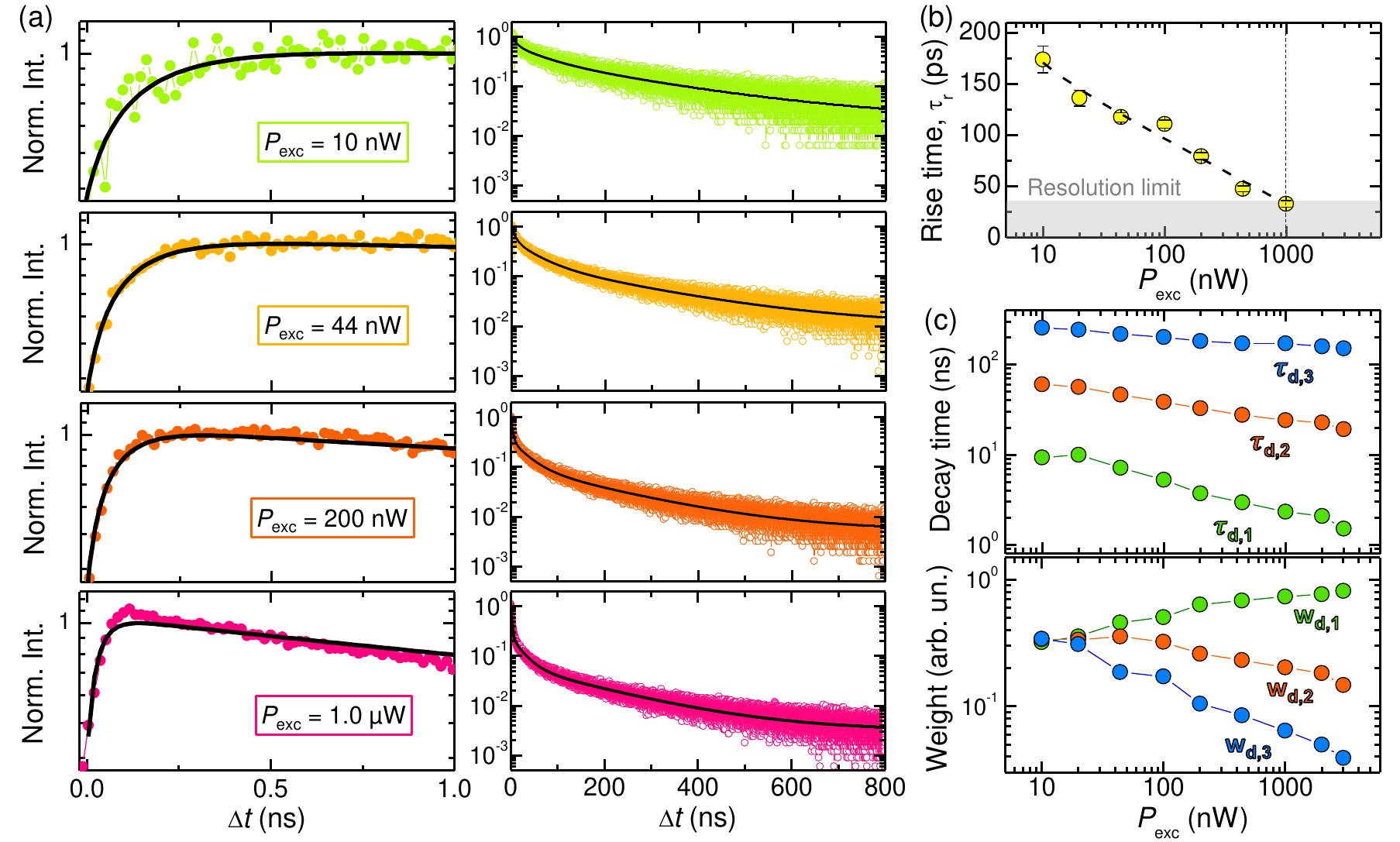}
		\caption{(a) Some exemplifying plots of the time-evolution of the $\mu$-PL signal of the investigated WSe$_2$/MoSe$_2$ HS recorded at 6 K for different excitation powers $P_\mathrm{exc}$, in the $\Delta t$=0-800 ns interval from the laser pulse. The detection energy was set at the MX-IX band. The solid lines are fit to the data by Eqs.\ (2) (left column) and (1) (right column) of the main text.
	(b) Summary of the estimated rise times $\tau_\mathrm{r}$ as a function of $P_\mathrm{exc}$. Above 1 $\mu$m (as highlighted by the vertical dashed line) the rise time goes below the resolution limit of our setup. (c) Summary of the estimated decay times $\tau_\mathrm{d,n}$ as a function of $P_\mathrm{exc}$ and corresponding weights $w_\mathrm{d,n}$.}
		\label{fig:PL_vs_P}
	\end{figure}

\begin{figure}[!ht]
		\centering
		\includegraphics[width=9cm]{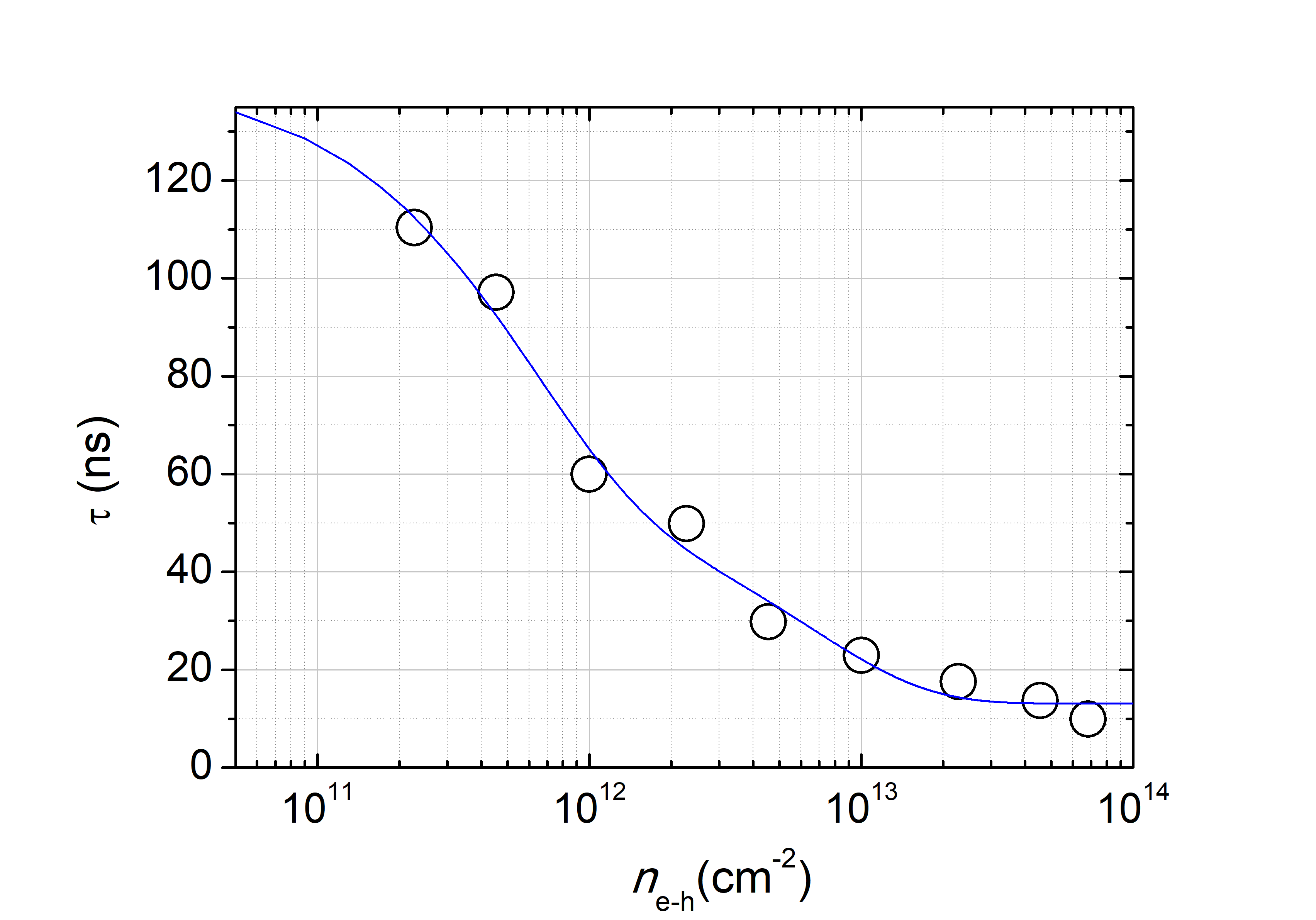}
		\caption{Dependence of the weighted decay time $\tau$ (symbols; see Table \ref{tab:tau}) on the density of photogenerated carriers $n_\mathrm{e-h}$. The solid line is a bi-exponential fit used to interpolate the data.}
		\label{fig:tau_vs_n}
	\end{figure}

	$\tau$($n_\mathrm{e-h}$) was deduced in two steps. We first measured the time decay of the exciton $\mu$-PL signal, whose temporal traces are shown in Fig. \ref{fig:PL_vs_P} for different $P_\mathrm{exc}$s. We then fitted the data using Eq. (1) in the main text with $n=3$ components characterised by their decay time $\tau_\mathrm{d,n}$ with relative weight $w_\mathrm{d,n}$. Table \ref{tab:tau} reports the $\tau_\mathrm{d,n}$ and $w_\mathrm{d,n}$ values for different $P_\mathrm{exc}$s along with the weighted value of the decay time $\tau$. Finally, we obtained $\tau$($n_\mathrm{e-h}$) using the following relationship under pulsed excitation between the injected carrier density \emph{n} and $P_\mathrm{exc}$:
	\begin{equation}
        n = \frac{P_\mathrm{exc} \cdot \sigma}{S \cdot f_\mathrm{rep} \cdot h\nu},
        \label{n0}
    \end{equation}
	where $\sigma$=0.08 is the absorbance of the MoSe$_2$/WSe$_2$ HS evaluated for h$\nu$=2.33 eV as reported in Ref. \cite{Wang_moire_mott} of this Supplementary Material, and $f_\mathrm{rep}$=1.2 MHz is the pulsed laser repetition rate. Therefore, we can relate \emph{n} to $P_\mathrm{exc}$ and hence to $\tau$ via Table \ref{tab:tau}. Fig. \ref{fig:tau_vs_n} shows the dependence of $\tau$ on \emph{n}.
	
\begin{table}[h!]
\caption{Fitting parameter values shown in Fig. \ref{fig:PL_vs_P}(b). $\tau_\mathrm{d,n}$ is the decay time relative to the $n$-th component, whose weight is given by $w_\mathrm{d,n}$. The last column reports the weighted decay time $\tau$.}
\begin{center}
{\renewcommand{\arraystretch}{1.2}
\begin{tabular*}{0.75\columnwidth}{@{\extracolsep{\fill}}cccccccc@{}}
\hline
\hline
$P_\mathrm{exc}\;(\mathrm{nW})$ & $\tau_{\mathrm{d,1}}\;(\mathrm{ns})$ & $w_{\mathrm{d,1}}$ &  $\tau_{\mathrm{d,2}}\;(\mathrm{ns})$ & $w_{\mathrm{d,2}}$ & $\tau_{\mathrm{d,3}}\;(\mathrm{ns})$ & $w_{\mathrm{d,3}}$ & $\tau\;(\mathrm{ns})$\tabularnewline
\hline
10 & 9.3 & 0.32 & 60.2 & 0.34 & 254.9 & 0.34 & 110.4\tabularnewline
20 & 10 & 0.36 & 56.2 & 0.34 & 242.0 & 0.31 & 97.1\tabularnewline
44 & 7.2 & 0.46 & 46.3 & 0.36 & 216.4 & 0.19 & 60.0\tabularnewline
100 & 5.3 & 0.50 & 38.4 & 0.32 & 201.2 & 0.17 & 49.9\tabularnewline
200 & 3.7 & 0.64 & 32.7 & 0.26 & 180.9 & 0.10 & 29.8\tabularnewline
440 & 3.0 & 0.68 & 27.5 & 0.23 & 171.4 & 0.08 & 22.9\tabularnewline
1000 & 2.3 & 0.73 & 24.1 & 0.20 & 170.8 & 0.06 & 17.6\tabularnewline
2000 & 2.1 & 0.77 & 22.6 & 0.18 & 159.2 & 0.05 & 13.7\tabularnewline
3000 & 1.5 & 0.81 & 19.1 & 0.15 & 150.7 & 0.04 & 9.9\tabularnewline
\hline
\end{tabular*}}
\end{center}
\label{tab:tau}
\end{table}
	
\begin{figure}[!ht]
		\centering
		\includegraphics[width=12cm]{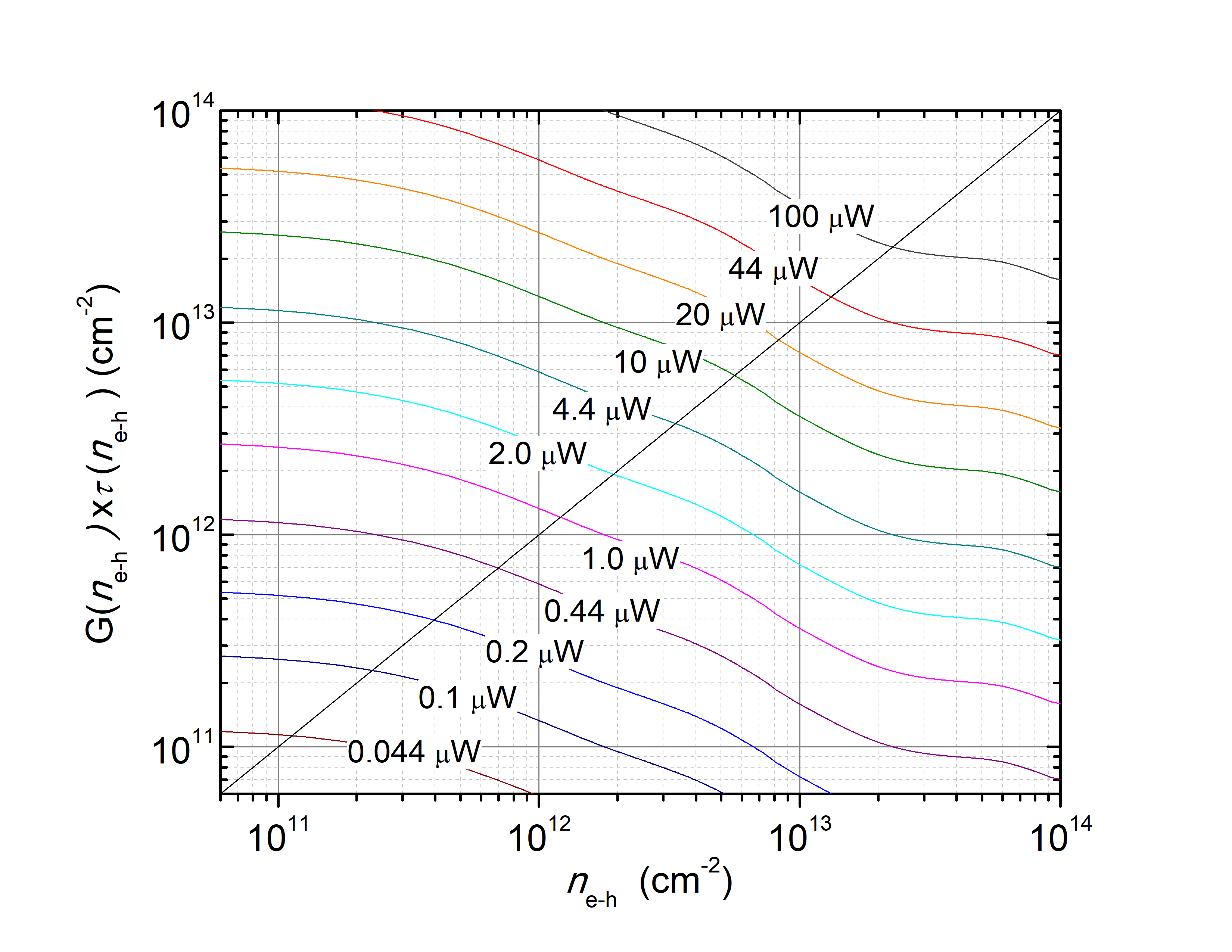}
		\caption{Plot of $G$($n_\mathrm{e-h}$) $\cdot$ $\tau$($n_\mathrm{e-h}$) \emph{vs} $n_\mathrm{e-h}$. The intersections between the bisecting black line and the other curves gives the solutions of Eq. \ref{n_vs_t} for each of the cw $P_\mathrm{exc}$ values considered in Fig. 3(a) of the main text.}
		\label{fig:g_dot_n}
	\end{figure}

	Given the  results shown in Figs. \ref{fig:absorbance} and \ref{fig:tau_vs_n}, we can solve Eq. \ref{fig:tau_vs_n} --namely $n_\mathrm{e-h}$=$G$($n_\mathrm{e-h}$) $\cdot$ $\tau$($n_\mathrm{e-h}$)--  numerically as displayed in Fig. \ref{fig:g_dot_n}. The solutions are given by the crossing of the first and second member of the above equation plotted on the \emph{y} and \emph{x} axis, respectively. That plot finally allows us to give the density of photogenerated electron-hole pairs for a specific excitation power $P_\mathrm{exc}$.

	 \clearpage
	 
	\newpage
	\subsection*{Supporting Note 4. Integrated photoluminescence intensity \emph{vs} power density for different temperatures}
	\addcontentsline{toc}{section}{Supporting Note 4. Integrated photoluminescence intensity \emph{vs} power density for different temperatures}
	\setcounter{section}{4}
        \sectionmark{}
	\setcounter{figure}{0}
        \setcounter{equation}{0}

\begin{figure}[!ht]
		\centering
		\includegraphics[width=16cm]{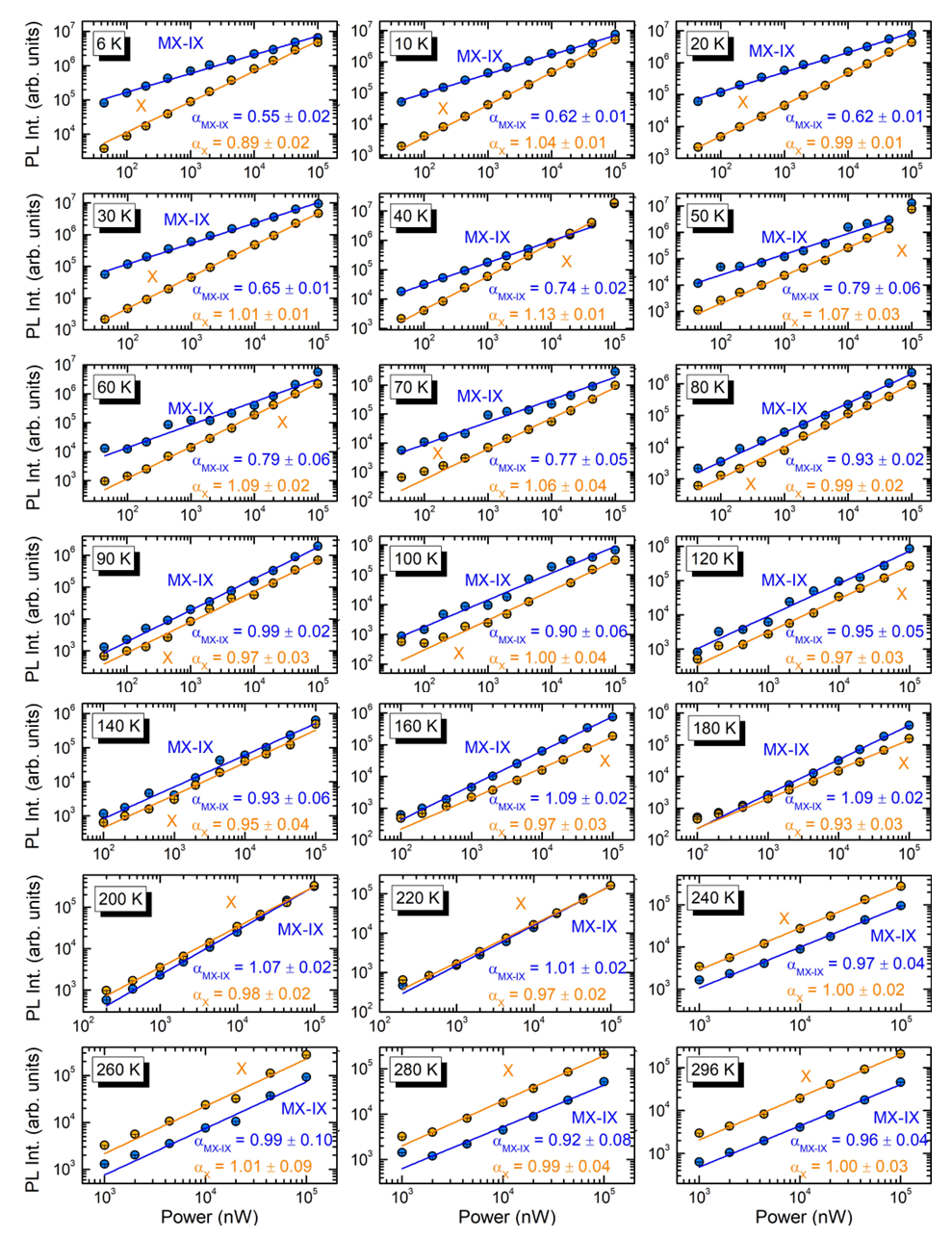}
		\caption{PL integrated intensity dependence on the laser power for MX-IX (azure symbols) and X (dark yellow symbols) bands at different temperatures. Solid lines are fits to the data via Eq. \ref{eq:IvsP_supp}. The $\alpha$ coefficient values obtained from the fits are displayed within each plot.}
		\label{fig:Optical-vs-AFM}
	\end{figure}

	The data shown in Fig. 3(c) of the main text were derived by fitting the photoluminescence integrated intensity \emph{I} at different temperatures using the following equation\
	
	\begin{equation}
    I = A\cdot{P_\mathrm{exc} ^{~\alpha}},
    \label{eq:IvsP_supp}
    \end{equation}
where $P_\mathrm{exc}$ is the laser excitation power, $A$ is a scaling constant and $\alpha$ a coefficient, whose value suggests the type of transition involved (\emph{e.g} $\alpha$=1 exciton transition, $\alpha$=2 uncorrelated electron-hole pairs, $\alpha$<1 finite density two-level system). Figure \ref{fig:Optical-vs-AFM} shows the dependence of \emph{I} on $P_\mathrm{exc}$ for the two main bands observed MX-IX (interlayer exciton, either moir\'{e}, MX, or free, IX) and X (free exciton of the heterostructure constituents); see, \emph{e.g.} Figs 2 (a) and 3 (a) in the main text. The measurements were performed at differnt temperatures between 6 K and 296 K. The results of the fits are displayed in the different panels of Figure \ref{fig:Optical-vs-AFM}.
	
	
	
	\clearpage
	
	\newpage
	\subsection*{Supporting Note 5. Temperature-dependent micro-photoluminescence}
	\addcontentsline{toc}{section}{Supporting Note 5. Temperature-dependent micro-photoluminescence}
	\setcounter{section}{5}
        \sectionmark{}
	\setcounter{figure}{0}
	
\begin{figure}[!ht]
		\centering
		\includegraphics[width=14cm]{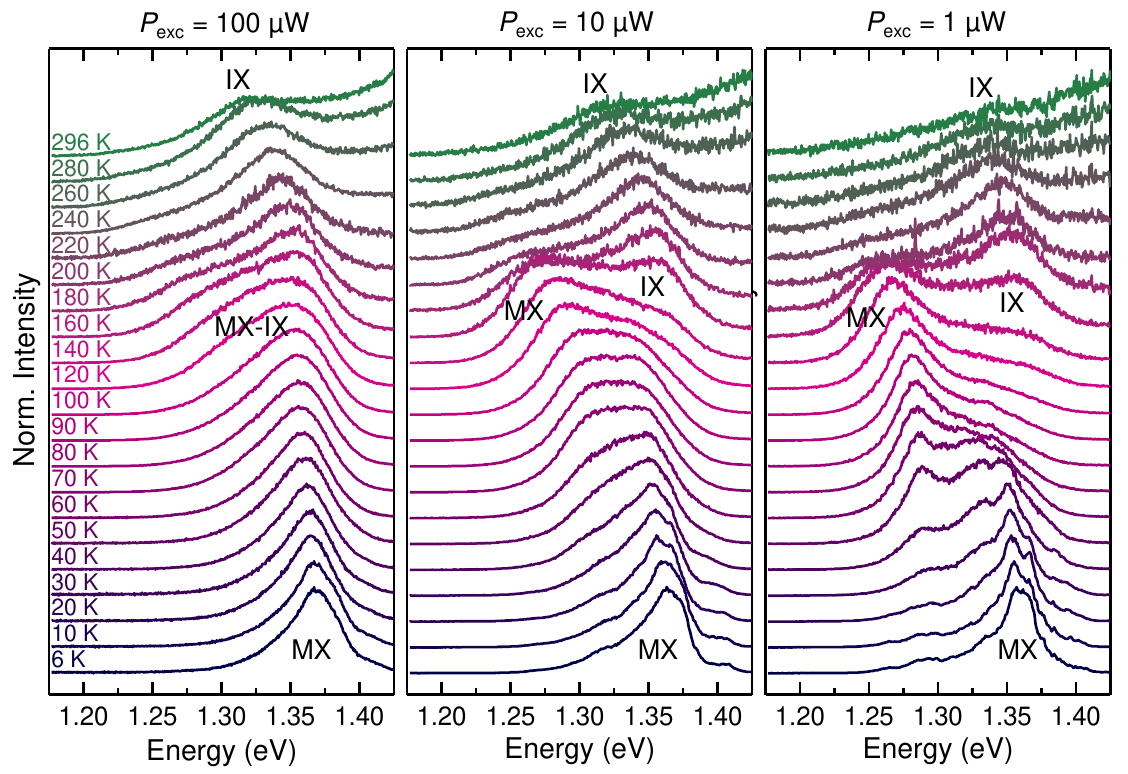}
		\caption{$\mu$-PL spectra $vs$ $T$ at three different laser excitation powers $P_\mathrm{exc}$ (focused via a 20$\times$ objective with NA = 0.4). MX and IX indicate the recombination band due to moir\'{e}-localised excitons (MXs) and free interlayer excitons (IXs), respectively.}
		\label{fig:trPL-vs-T}
	\end{figure}

	Figure 4(a) in the main text shows the temperature dependence of the $\mu$-photoluminescence (PL) spectra recorded on the investigated WSe$_2$/MoSe$_2$ heterostructure for a given laser excitation power $P_\mathrm{exc}$ (=10 $\mu$W). The set of data shows a clear \emph{T}-induced variation in the emission lineshape caused by the de-trapping of moir\'{e}-localised excitons (MXs) in favour of free interlayer excitons (IXs). Figure \ref{fig:trPL-vs-T} shows a similar study performed at a higher (=100 $\mu$W) and lower (=1 $\mu$W) $P_\mathrm{exc}$. Figure \ref{fig:trPL-vs-T} indicates that, for a given \emph{T}, the de-trapping process becomes more apparent for a larger density of photogenerated carriers (\emph{i.e.} larger $P_\mathrm{exc}$).
  	
\begin{figure}[ht!]
		\centering
		\includegraphics[width=7cm]{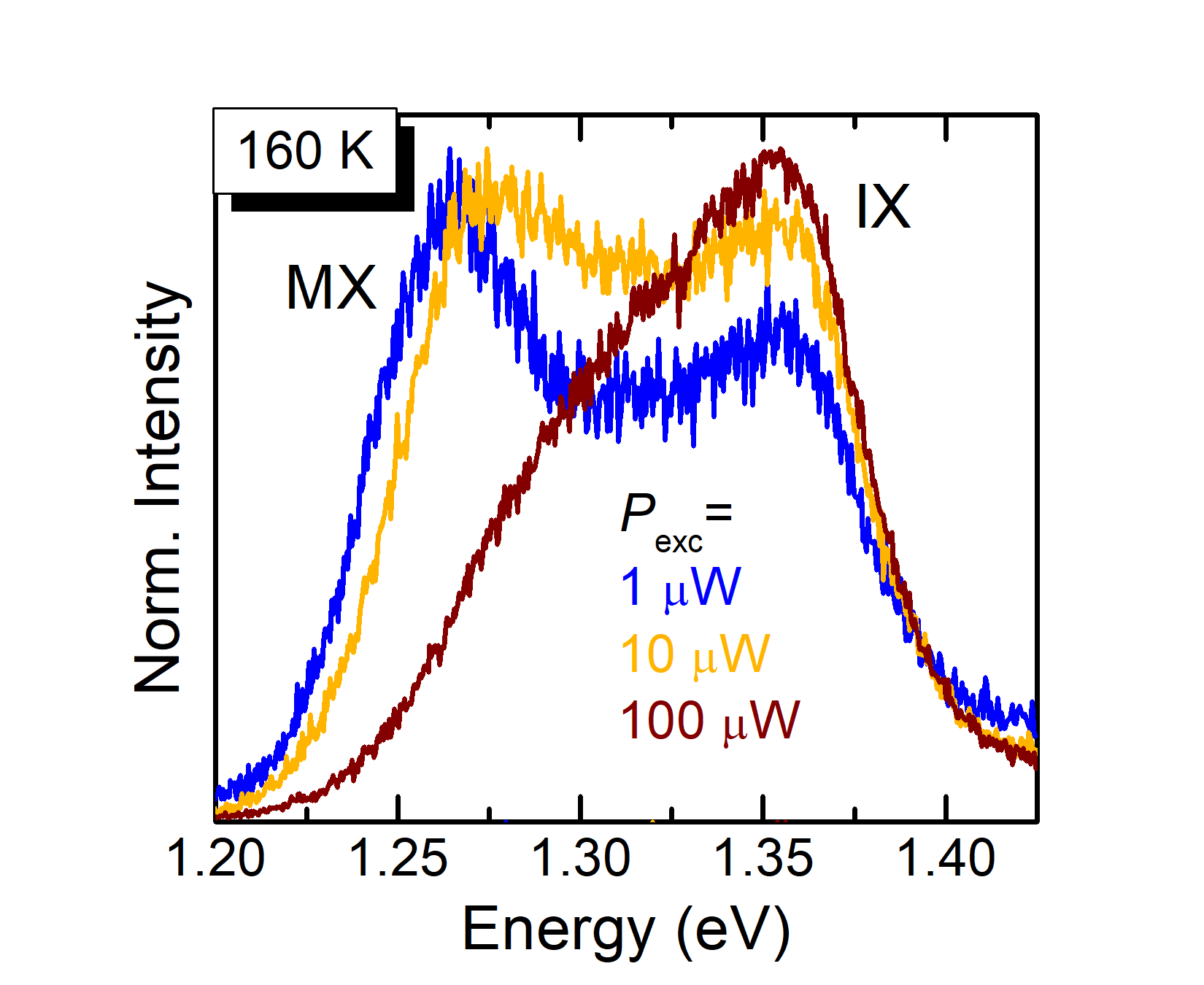}
		\caption{$\mu$-PL spectra at 
        160 K
        for different laser excitation powers $P_\mathrm{exc}$ (focused via a 20$\times$ objective with NA = 0.4). MX and IX indicate the recombination band due to moir\'{e}-localised excitons (MXs) and free interlayer excitons (IXs), respectively.}
		\label{PL_vs_P_Tfixed}
	\end{figure}

 This can be better appreciated in Figure \ref{PL_vs_P_Tfixed}. Indeed, for fixed \emph{T}=160 K, the relative weight of the IX component increases for increasing number of photogenerated carriers and for the ensuing saturation of the finite moir\'{e} potential sites. This saturation takes place more evidently as \emph{T} increases, which favours moir\'{e} de-trapping as shown in the previous figure.

\clearpage
	
	\newpage
	\subsection*{Supporting Note 6. Temperature-dependent $g$-factor of the MX/IX band}
	\addcontentsline{toc}{section}{Supporting Note 6. Temperature-dependent $g$-factor of the MX/IX band}
	\setcounter{section}{6}
        \sectionmark{}
	\setcounter{figure}{0}
        \setcounter{equation}{0}
	
\begin{figure}[!ht]
		\centering
		\includegraphics[width=1.0\textwidth]{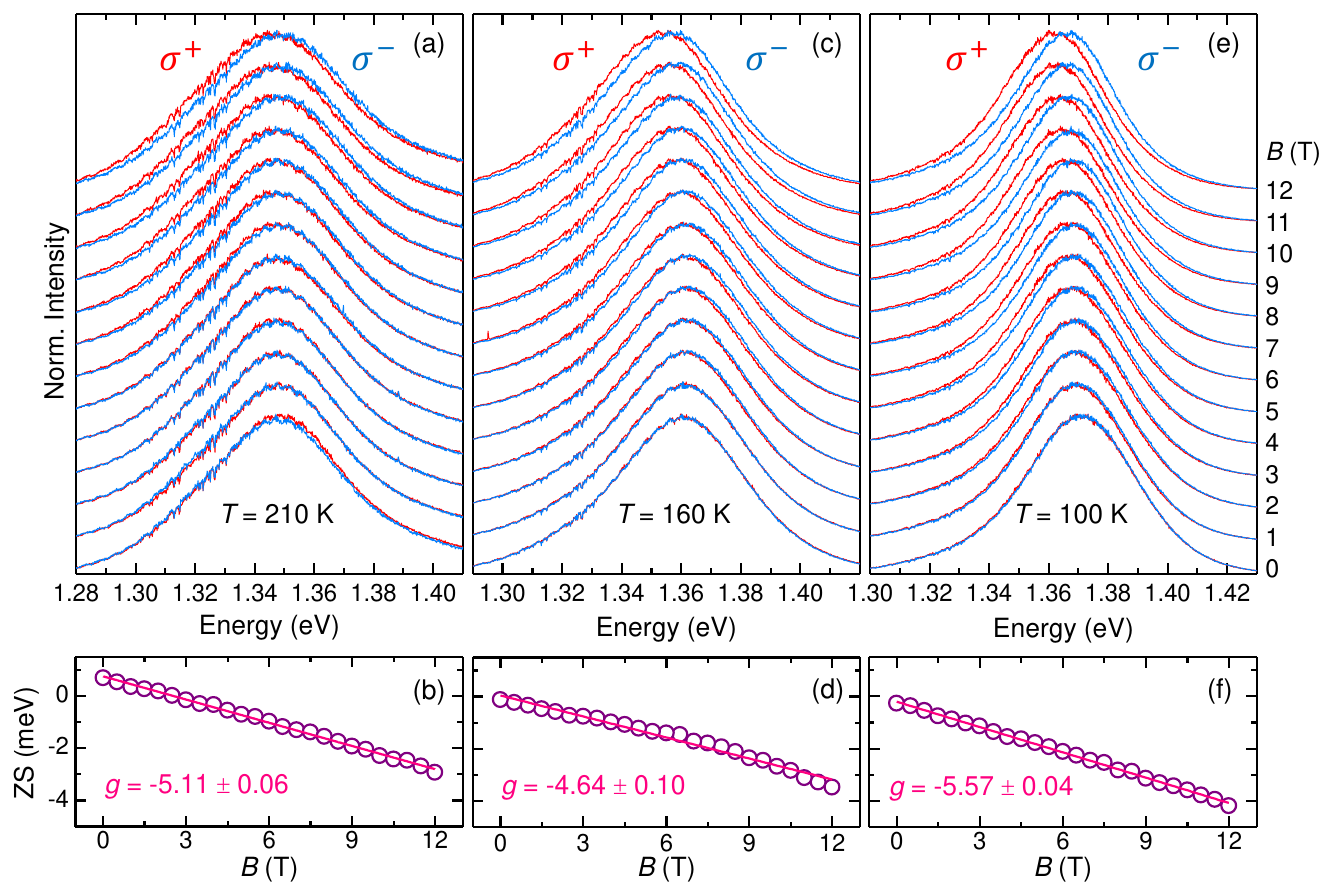}
		\caption{(a) Helicity-resolved normalised $\mu$-PL spectra $vs$ magnetic field at $T = 210$ K. (b) ZS of the band, from which the displayed $g$-factor was obtained through a linear fit. (c)-(d) Same as panels (a)-(b) for $T = 160$ K. (e)-(f) Same as panels (a)-(b) for $T = 100$ K. 
        It should be noticed that the standard deviation associated to each fit clearly underestimates the real error on the $g$-factor. The latter is in fact a bit dependent on the exact sample position where the measurements are taken. By staying at the maximum field (12 T) and changing position, we observed ZS variations within 20-25 $\%$. The results of panels (b), (d) and (f) thus show that within the position-related uncertainty, a similar $g$-factor of $\approx + 5$ is found at elevated temperatures.
        The same excitation power $P_\mathrm{exc} = 75 ~\mu$W (focused via a 100$\times$ objective with NA = 0.8) was used for all the sets of data.
            }
		\label{fig:g-factor_vs_T}
	\end{figure}

	Figure \ref{fig:g-factor_vs_T} shows a series of $\mu$-photoluminescence (PL) spectra recorded at different magnetic fields for \emph{T}=210, 160 and 100 K, panels (a), (c) and (e), respectively. The spectra were recorded with opposite circular polarisation filtering ($\sigma^+$ and $\sigma^-$) and the Zeeman splitting (ZS) was then derived. The ZS dependence on \emph{B} is shown in panels (b), (d) and (f) for \emph{T}=210, 160 and 100 K, respectively. The data were fitted by
\begin{equation}
\mathrm{ZS}(B) = E^{\sigma^+} - E^{\sigma^-} = g \cdot \mu_\mathrm{B} B,
\label{eq:ZS_supp}
\end{equation}
where $E^{\mathrm{\sigma^{\pm}}}$ are the peak energies of components with opposite helicity $\sigma^+$ and $\sigma^-$, $g$ is the exciton $g$-factor, and $\mu_\mathrm{B}$ is the Bohr magneton. At the considered temperatures, $g$ is negative and corresponds to the gyromagnetic factor of the free interlayer exciton $g_{\mathrm{exc},\mathrm{IX}}$. The values found for different \emph{T}s are shown in the corresponding panels and do not change appreciably with temperature. As discussed in the main text, the $g_{\mathrm{exc},\mathrm{IX}}$ sign and absolute value can be ascribed to the avoided action of the moir\'{e} potential determined by the \emph{T}-induced MX de-trapping. Indeed, $g_{\mathrm{exc},\mathrm{IX}}$ can be estimated by considering the separate contribution
of electrons and holes to the IX gyromagnetic factor (see Eq. (5) in the main text).
  	
 \clearpage
	
	\newpage
	\subsection*{Supporting Note 7. Power-dependent $g$-factor of the MX/IX band at low $T$}
	\addcontentsline{toc}{section}{Supporting Note 7. Power-dependent $g$-factor of the MX/IX band at low $T$}
	\setcounter{section}{7}
        \sectionmark{}
	\setcounter{figure}{0}
	
\begin{figure}[!ht]
		\centering
		\includegraphics[width=0.75\textwidth]{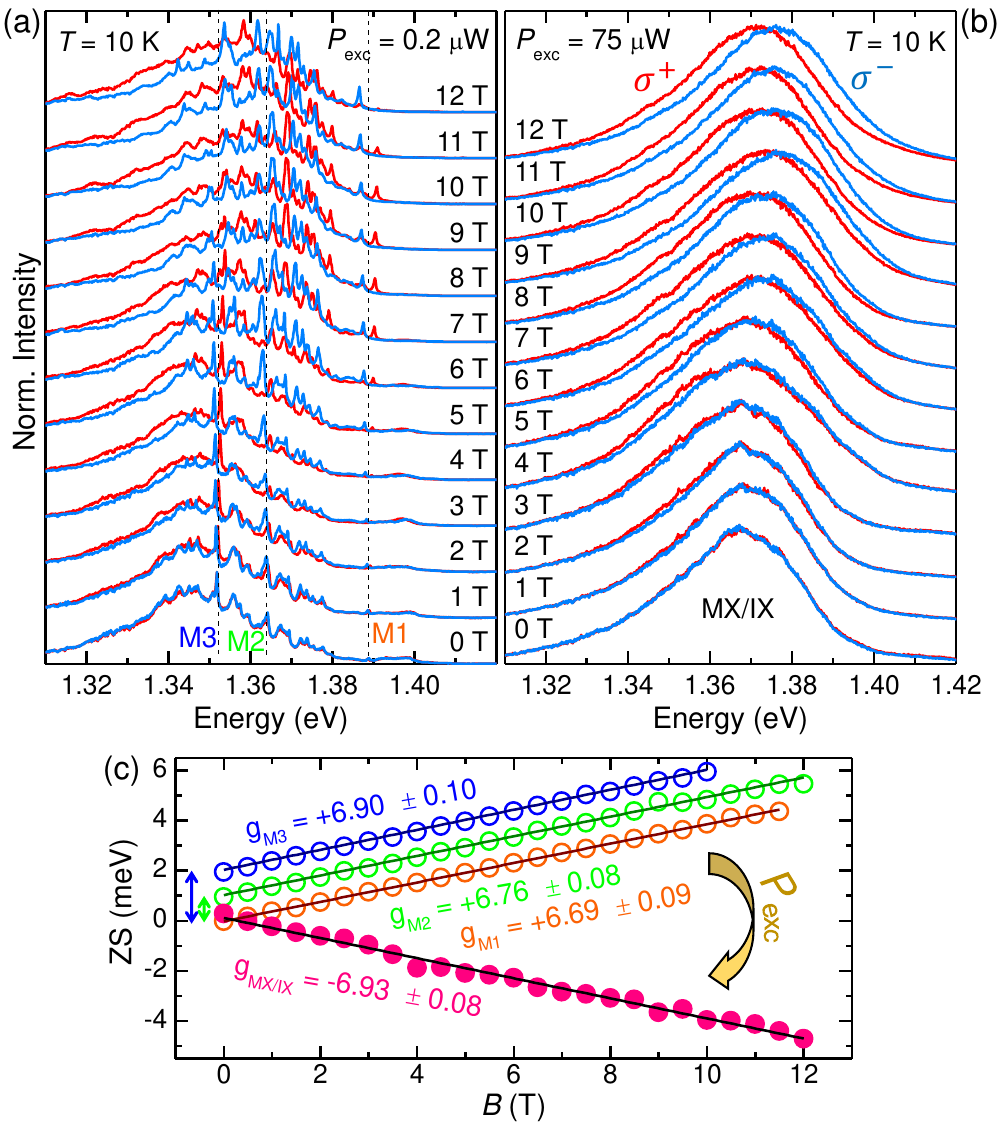}
		\caption{ (a-b) Helicity-resolved (red lines corresponding to the $\sigma^+$ polarisation, and blue lines to the $\sigma^-$ one) normalised $\mu$-PL spectra under magnetic field at $T = 6$ K for two different laser excitation powers $P_\mathrm{exc}$ (focused via a 100$\times$ objective with NA = 0.75). The two sets of data were acquired in the same point of the HS.
        For $P_\mathrm{exc} = 0.2 ~\mu$W (panel (a)), many narrow lines can be seen. M1, M2 and M3 indicate three such narrow lines (M1 corresponds to line M1 in Fig.\ 4 of the main text). At high powers $P_\mathrm{exc} = 75 ~\mu$W (panel (b)) a continuous band can be seen. (c) ZS of the lines M1-M3 of panel (a) and of the MX/IX band of panel (b), showing an opposite sign of the $g$-factor. The data of lines M2 and M3 are up-shifted by 1 and 2 meV, respectively (as indicated by the double-sided arrows on the left) for sake of clarity.
            }
		\label{fig:g-factor_low_T_vs_P}
	\end{figure}

	Figure \ref{fig:g-factor_low_T_vs_P}(a) shows a series of $\mu$-photoluminescence (PL) spectra recorded at different magnetic fields for \emph{T}=6 K and low laser excitation power $P_\mathrm{exc}$=0.2 $\mu$W. Several narrow lines (denoted as ''Mi'') due to moir\'{e} confined excitons can be observed, superimposed on a continuum background. The spectra were recorded with opposite circular polarisation ($\sigma^+$ and $\sigma^-$) with each narrow line exhibiting a positive Zeeman splitting $\mathrm{ZS}(B) = E^{\sigma^+} - E^{\sigma^-} = g_\mathrm{Mi} \cdot \mu_\mathrm{B} B$. $E^{\mathrm{\sigma^{\pm}}}$ are the peak energies of components with opposite helicity $\sigma^+$ and $\sigma^-$, $\mu_\mathrm{B}$ is the Bohr magneton and $g_\mathrm{Mi}$ is the exciton gyromagnetic factor pertaining to the i-th moir\'{e} confined exciton.
 Panel (b) shows that increasing $P_\mathrm{exc}$ by about a factor 400 the narrow lines associated to moir\'{e} confined excitons merge in a continuum of states, whose maximum shows a sizeable blue-shift of more than 20 meV. As discussed in Fig.\ 3 of the main text, for high laser power the emission band comprises a mixture of free interlayer excitons IXs and moir\'{e} confined excitons. As a matter of fact, the ZS is negative, namely the $E^{\sigma^+}$ component is at lower energy with respect to the $E^{\sigma^-}$ component.
 Figure \ref{fig:g-factor_low_T_vs_P}(c) shows the ZS dependence on magnetic field for the three moir\'{e} confined excitons highlighted in panel (a), where $P_\mathrm{exc}$=0.2 $\mu$W, and for the continuum of states obtained for $P_\mathrm{exc}$=75 $\mu$W. The largely different gyromagnetic factors observed in the two cases are clear. Under high excitation conditions,  exciton-exciton interaction effects likely tend to screen the moir\'{e} potential and restore the electronic properties of the free IXs, whose gyromagnetic factor is characterised by a negative sign.
  	
 \clearpage

\newpage

\renewcommand{\bibname}{References}
\addcontentsline{toc}{section}{References}

\bibliographystyle{thesisamj}
\bibliography{jabbr,mybib}

\end{document}